\documentclass[fleqn,11pt]{wlscirep}
\usepackage{graphicx,epsfig}

\usepackage{graphics,dcolumn,bm,fleqn,epic,eepic}
\usepackage{amssymb,amsmath,multirow,rotate,color}
\usepackage[caption=false]{subfig}
\usepackage{color}
\usepackage{soul}
\usepackage{tikz}
\usetikzlibrary{patterns}
\usepackage{graphicx}

\usepackage[T1]{fontenc}

\usepackage{ulem}
\usepackage{listings,textcomp}

\usepackage[utf8]{inputenc}

\usepackage[english]{babel}
\usepackage{braket}

\definecolor{FI}{rgb}{1., 1.0,.2} 
\definecolor{JP}{rgb}{1, 0.5, 0.5} 
\definecolor{CT}{rgb}{0.5, 1.0, 0.83} 

\date{\today}

\title{Stochastic modelling of blockchain consensus}
\author[1,2,*]{Claudio J.~Tessone}
\author[3]{Paolo Tasca}
\author[2]{Flavio Iannelli}
\affil[1]{UZH Blockchain Center, Universit\"at Z\"urich, Andreasstrasse 15, CH-8050 Z\"urich, Switzerland}
\affil[2]{URPP Social Networks, Universit\"at Z\"urich, Andreasstrasse 15, CH-8050 Z\"urich, Switzerland}
\affil[3]{Centre for Blockchain Technologies, University College London, United Kingdom}

\affil[*]{claudio.tessone@uzh.ch}

\begin{abstract}

Blockchain and general purpose distributed ledgers are foundational technologies which bring significant innovation in the infrastructures and other underpinnings of our socio-economic systems.  
These P2P technologies are able to securely diffuse information within and across networks, without need for trustees or central authorities to enforce consensus. 
In this contribution, we propose a minimalistic stochastic model to understand the dynamics of blockchain-based consensus. 
By leveraging on random-walk theory, we model block propagation delay on different network topologies and provide a classification of blockchain systems in terms of two emergent properties.
Firstly, we identify two performing regimes: a functional regime corresponding to an optimal system function; and a non-functional regime characterised by a congested or branched state of sub-optimal blockchains.
Secondly, we discover a phase transition during the emergence of consensus and numerically investigate the corresponding critical point. 
Our results provide important insights into the consensus mechanism and sub-optimal states in decentralised systems.
 
\end{abstract}

\begin{document}
\maketitle

\section{Introduction}

Emerging techno-social systems composed of many interacting components are pervasive in our society \cite{vespignani2009predicting}. Prominent examples include the World Wide Web and the physical Internet, the inter-bank financial network, power distribution grids and the global mobility network \cite{pastor2007evolution,battiston2010structure,barrat2004architecture}.
Network science is the natural framework to study complex techno-social systems in combination with sophisticated computational models.
Among the many emerging dynamical self-organised systems, with the Internet as the primary example, in recent years distributed ledger technology is attracting the attention of the scientific community \cite{aste2017blockchain}. Our work resorts to complex network theory, which turns to be very useful for understanding distributed systems: on one side, it allows for their mathematical description in terms of purely topological properties. On the other side, with the use of computational agent-based models, it enables the modelling of their  collective behaviour.

In particular, the focus of our analysis is on blockchain systems. Generally included in the larger family of distributed ledger technologies \cite{tasca2019taxonomy,spychiger2020}, a blockchain is a sequence of blocks, each of them containing transaction data that is the logical continuation of the preceding one. 
Records of all the transactions ever occurred in the network are included in a public ledger, which is generally open for scrutiny.
Instead of relaying and being controlled by a central authority, every node - validator or miner -  connected to the blockchain peer-to-peer (P2P) network has a local replica (not a copy!) of the records and can validate new blocks of transactions. 
Thus, the system is characterised by history redundancy across the whole network, making it especially resistant to external perturbations, such as targeted attacks and node(s) failure.
Here, it is important to note that “validation” and “consensus” are not the same thing. A blockchain validator performs validation by verifying that transactions are legal (not malicious, double spends, etc). Instead, consensus is performed by miners, and involves determining the ordering of events in the blockchain and coming to agreement on that order. That agreement is reached either via Proof-of-Work (PoW) or Proof-of-Stake (PoS) protocols, which provide a probabilistic solution to the byzantine generals problem for consensus.\cite{swan2019blockchain}.

In our work, we present a minimalist modelling of blockchain consensus and analyse PoW and PoS-emergent properties using a complex systems approach. 
Consensus modelling is generally the result of both a cooperative process among the agents, and a competitive process with respect to the alternatives available to the agents
\cite{baronchelli2018emergence}.
In the case of blockchain systems, cooperation is driven by the diffusion of blocks among the P2P network nodes, and newly mined blocks represent the alternatives.

An important feature of blockchain-based systems is that consensus among the elements is emergent and not enforced. In our model, a set of nodes involved in the process of mining are connected through a P2P network.  Each node may discover a new block at a certain rate (proportional to its computational or staking power), and diffuse it by means of gossiping to other nodes in the network. Ideally, when the diffusion of blocks is fast enough, all blocks are chained one after the other. However, not all blocks consisting of completely valid transactions achieve consensus by the network.
Indeed, each time a transaction is made, it is broadcast to the entire network. Upon hearing the broadcasts, miners take a bunch of transactions, validate that they are ``legitimate'' and put them into a block.
But miners ``hear'' about different transactions at different times (due to latency issues, among others). Furthermore, they may simply choose different transactions to include in their block, based on the transaction fees. Miners do not need to build the same global block. They can each build their own blocks consisting of entirely different transactions. If multiple blocks are found before diffusion across the network, the blockchain turns into a \textit{blocktree}, whereby miners build new blocks on a different previous block and the blockchain splits.
In this memory-dependent process, therefore, there exists a latent, unresolved competition between block creation and block diffusion. Blocks that are not able to diffuse through the network are wasted resources.


The blocktree is a global, emergent property of the diffusion process. We show that its characterisation allows to assess the system efficiency. 
Our approach enables studying the path towards consensus depending on few meaningful variables. In this way, we can identify the conditions under which a blocktree emerges: the full set of blocks that are mined concurrently by different nodes, and the overall network characteristics.

The topology of the blocktree serves as an indicator of the efficiency of system design.
The more branches emerge, the more resources are wasted without contributing to the allocation of information in a distributed fashion.
By relaxing the assumption of direct path propagation of blocks, we find an accurate estimate of blockchain forking, in terms of the random-walk hitting time on the P2P network. 
In addition, we identify a phase transition to consensus that can be formalised in terms of an ``order'' parameter and a ``control'' parameter defined by the block diffusion time in the P2P network of miners. 
To conclude our analysis, we perform numerical simulations and compare our results with the observed data from different blockchain-based systems.


\section{The system}

\paragraph{Description of the system}



We first describe the system of non-permissioned blockchains (the most widely deployed ones in the market), and those that more naturally self-organise into a normal working state.

A block contains a set of atomic operations whose validity has been verified by the block creator. The most typical usage of blockchain-based systems is as underlying infrastructure for the so-called \textit{cryptocurrencies}; in this context, the atomic operations are characterised by transactions between system participants \cite{tasca2015digital}. 
Their size is generally very small and, without loss of generality, we assume that its effect on system performance can be neglected. Therefore, we do not include block size in our modelling.

In order to disincentivize network participants from adding forged information into the system or acting maliciously, the process of adding information must be costly for the producer -- but relatively inexpensive for all others -- to verify its correctness. This premise is the celebrated solution to the ``Bizantine fault tolerance'' problem, and different systems implement the solution in different ways.  The earliest and most common one is PoW. In this solution, beyond validating all the transactions that will be included into the block, the miners need to solve a ``computational puzzle'' namely find the problem of inverting a cryptographic function which involves information relative to the previous block and the transactions contained in the current one. Thus, each block data is linked backward in a chain of blocks.

It is worth noting  that 
real-world blockchain-based systems adjust the difficulty of block discovery (i.e., the expected number of trials necessary to solve the ``computational puzzle'') with the aim of achieving a constant rate addition of new blocks to the blockchain, irrespective of the total computational power employed in the mining process. 
For example, in Bitcoin and its derivatives, the difficulty level is adjusted every two weeks, such that, on average, a new block gets produced every 10 minutes. On the other hand, Ethereum adjusts its difficulty level so that the expected time for a new block to be produced is just 12 seconds.
In a short time interval, the probability that a miner finds the block is proportional to the number of attempts it has made to find the solution to the ``computational puzzle'', with respect to the total number of attempts performed by all miners in the same time interval. 
In the specific case of Bitcoin, this created an arms race to devote more and more computational power.\footnote{Indeed the number of trials 
 to find a new block escalated $10^7$ times in the last 8 years.}

The winner of the ``computational puzzle'' competition (i.e., the creator of the last block) includes a new transaction  
(coinbase) -- without issuer address and with theirself as recipient--. For PoW-based cryptocurrency systems, this is the only way in which new currency enters in circulation, and standard practice is that the coinbase amount -- although it may change over time - is pre-determined. 

For the purpose of our analysis, it is important to distinguish between validity and consensus. With respect to the validity of the information contained in the blocks, only after a block has been accepted by all network participants, i.e.~they have reached consensus about it, it can be  assumed that the transactions contained therein have really taken place. In rare cases, two nodes may solve the ``computational puzzle'' and be selected to add a new block to the chain almost at the same time. Each of these nodes may choose a different next block with different transactions. They start to broadcast their version of the blockchain to the neighbouring nodes. In this case, we observe an inconsistency (fork) issue in the network, where different blocks have a different version of the blockchain. This problem can be solved with the ``longest chain'' principle, which states that the longest blockchain version is the valid one.

For a blockchain, its \textit{height} is the longest path that originates in the first block in the sequence.\footnote{The block \textit{height} of a particular block is defined as the number of blocks preceding it in the blockchain.}
Once a node finds a block, it broadcasts it to the list of peers that are known to it. Those peers will accept this block (and all intermediary ones not know yet to it) if its \textit{height} is larger than their local ledger replica.  In PoW and PoS, when two versions of the ledger have the same length, other nodes will be selected to add a new block on one of these ledger versions. The probability that the consensus mechanism chooses new blocks on different versions at the same rate becomes lower in time. This means that one of the variants will become predominant with more blocks and, due to the ``longest chain principle'', it will also become the valid one.



In our work we explore the space of parameters and find the conditions under which a PoW/PoS blockchain reaches consensus efficiently and study the phase transition point that leads to the emergence of forks.

\paragraph{The mechanics of the blockchain.}
In order to better appreciate the next section, Section \ref{sec:modelling}, we now briefly describe how PoW/PoS consensus protocols work when computing devices are interconnected through a logical P2P network. In this setting, peer nodes are assigned to each new node joining the network by master nodes that keep track of all nodes connected to the network

In principle, the peers assigned to a new entrant node are randomly drawn from a list of nodes assumed to be connected to the network.
Such selection is uniform (i.e. without favouring any node characteristics), in an attempt to make the network as unstructured as possible. However, the time distribution of node connectivity is very broad and highly skewed. Therefore, this naïve approach may induce a bias favouring nodes with long-time connectivity to gain more peers than the other ones, which either remain connected to the network for shorter periods of time or change their IP address.
As we explained earlier, every node (\textit{i}) hosts a local version of the blockchain. This is  composed of all the blocks transmitted to and accepted by \textit{i}  or blocks that \textit{i} generated by itself.\footnote{A node accepts a new block \textit{if} its \textit{height} parameter is bigger than that highest block the  node already has in its local replica of the ledger.} Node \textit{i} must also communicate the status of its blockchain to other peers that are known to \textit{i}. 

\begin{figure}
\centering
 \includegraphics[width=0.24\textwidth]{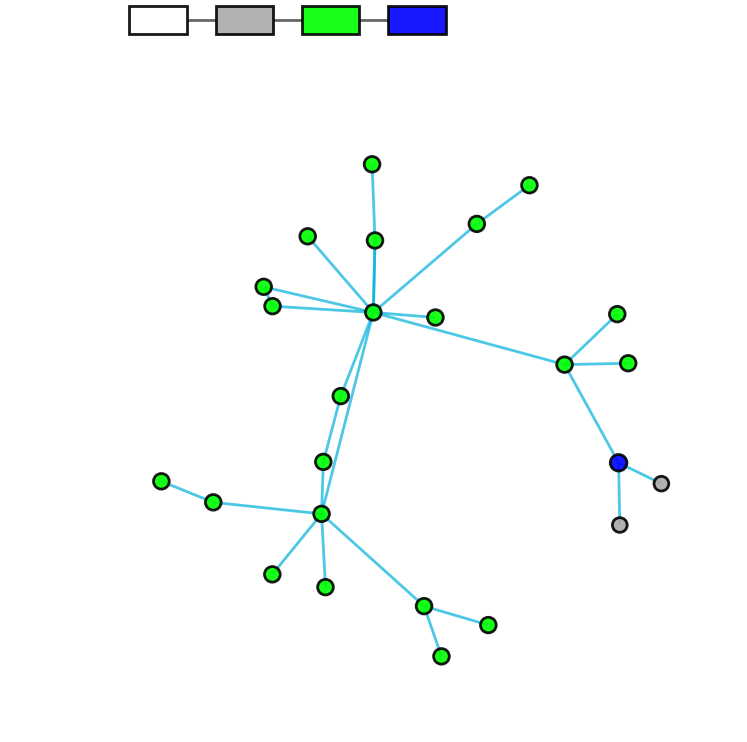}
 \includegraphics[width=0.24\textwidth]{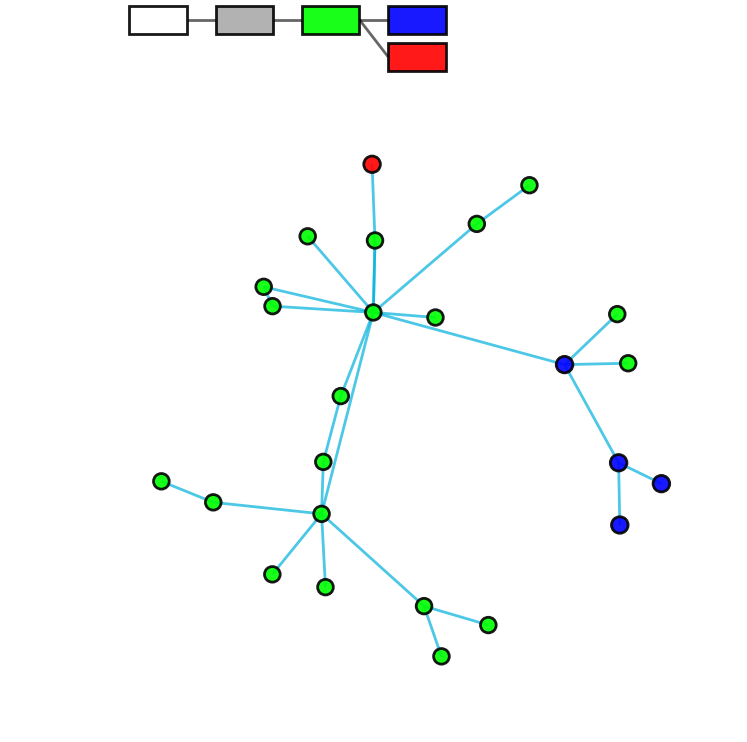}
 \includegraphics[width=0.24\textwidth]{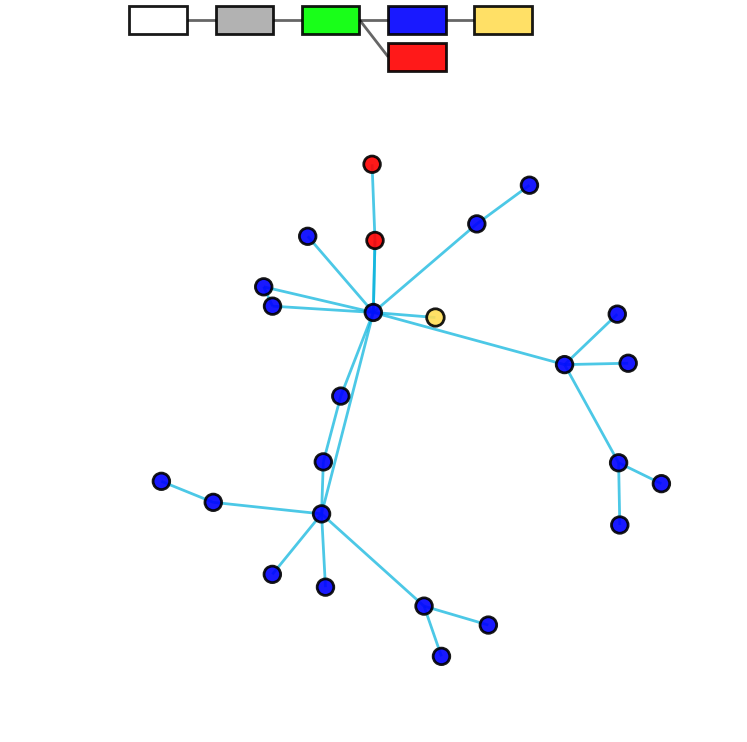}
 \includegraphics[width=0.24\textwidth]{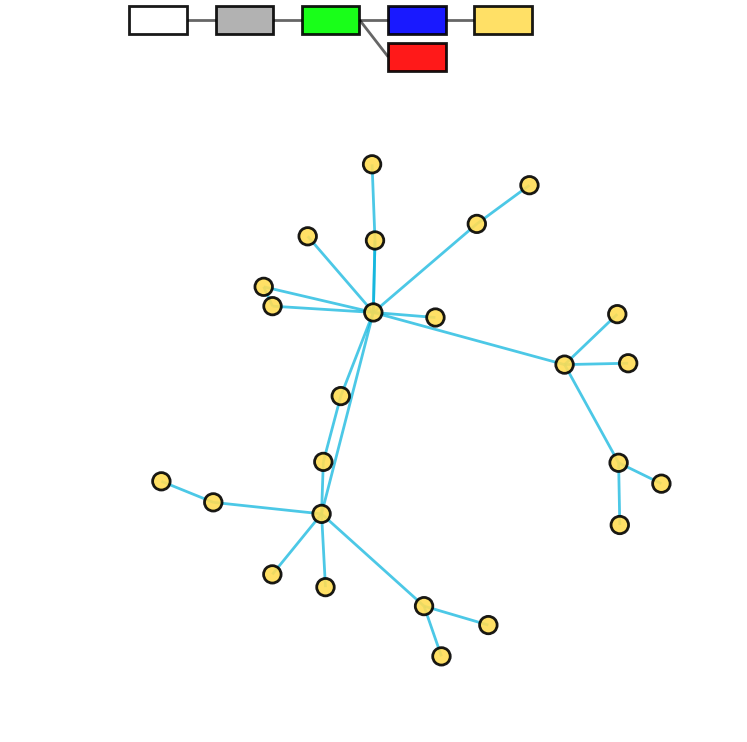}
 \caption{\label{fig:consensus}
Depiction of competing block diffusion in the P2P network and the corresponding  blockchain with nodes colour representing the highest block known to it.
}
\end{figure}

Let us see how this works in practice with a graphical example. In Fig.~\ref{fig:consensus} (leftmost panel), each block is represented by a different colour, and the colour of the highest block -- known to each node -- is taken by the nodes. In this example height $0 = white$, height $1 = grey$, height $2 = green$, height $3a = blue$. At time $t=0$ the majority of nodes is green with height $= 2$. The blue node has just mined a new block that has not yet been propagated in the network. The centre-left panel shows an intermediate situation $t=1$ where the blue node propagates to its neighbour nodes the blue block with height $=3a$ and they append it to their chain. However, at the very same time, before the blue node is able to reach all the network with the blue block, another node in the system finds an alternative block (coloured in red with height $=3b$) based on the green one of height $=2$.

In the centre-right panel, at time $t=2$, the blocks continue to propagate throughout the network. To notice that, when two neighbouring nodes have blocks of the same height, but the blocks are different (in our example, either blue or red), none of them accepts the block offered by the other one. At the same time $t=2$, a node has just discovered a new block of height $=4a$ based on the blue one of height $=3a$. The spreading of this new block is initiated throughout the network. Obviously, nodes whose longest chain ends in the blue node accept the new state. But now, blocks that have the red block as the highest must accept both: on the one hand they must include the blue block in their local blockchain and add the yellow one as well. Effectively, the red block is discarded from the history and the coinbase transactions are forever lost. However, the other transactions will be  imported again by the miners in their new pool of transactions.

After this example, we may observe some familiarity between blockchain consensus processes and epidemic ones: in absence of competing blocks, the evolution of the system resembles that of the celebrated susceptible-infected model \cite{pastor2015epidemic}. In presence of competing blocks, the spreading becomes a new kind of spreading process with memory: where the susceptibility of a node to receive the new state depends on whether this new state is acceptable (because of the height precedence rule) or not. This insight is key to introduce the model in Section \ref{sec:modelling}. 

The processes of block creation, spreading and eventual competition lead to emergent structures in the system that are non-trivial.
In principle, the systems are designed to lead to the creation of blockchains. However, the analytical derivation of their correct functioning is based on two important assumptions: full connectivity between network nodes, and negligible block propagation times. 
None of them are conditions that are satisfied as the system grows: on the one hand, because the network's diameter grows; on the other, because with increased user adoption, block size growth leads to non-vanishing propagation time. 

In absence of a simple mathematical derivation, we resort to a minimalistic modelling approach that helps us to study the impact of any change of blockchain technological parameters and concomitant protocols, and to analyse the behaviour of network participants, system growth and other dynamics.



%
%
%
%

\section{Modelling}
\label{sec:modelling}

\paragraph{Peer-to-peer network.} We consider a set of nodes connected to the P2P network. These nodes are involved in the process of transaction verification and -- as explained before -- block generation. 
Just for the sake of simplicity, we neglect the dynamics by which nodes enter and exit the network, given that the processes of node churn and renewal have characteristic time-scales, which are much longer than those of inter-block time and consensus  \cite{gencer2018decentralization}. 
We further assume that we can neglect changes on network topology induced by new peer addition. Thus, the network is defined by the number of nodes $N$ and the number of edges $E$. In existing protocols \cite{gencer2018decentralization,geier2019using}, the number of neighbours given to any new node is fixed. Therefore, another important parameter is the average degree $\braket{k}=2E/N$. In the simulations, we will consider different network topologies for the logical network and vary $\braket{k}$.

\paragraph{Node characteristics.} Each node connected to the network is endowed with some internal characteristic that ultimately determines its probability of creating a new block. In the case of PoW blockchains, it is the computational power -- measuring the number of trials it can make to solve the cryptographic problem -- $\pi_i$. Instead, in PoS systems, $\pi_i$ represents the total amount of assets staked for block creation. 
Importantly, the computational powers and stake allocations ($\pi_i$) are heterogeneous in the real consensus networks, and empirical analysis suggests very broad distributions \cite{li2020mining,li2020proof}. To study possible scenarios, we considered exponential and power-law distributions of computational power in the network nodes.

In a very small time interval $dt$, the probability that a specific node mines a new block is 
$
p^m_i \approx \eta_i dt = {\pi_i}/{ \sum_{\ell} \pi_{\ell} } . 
$
Since the events are independent and occur at fixed rate $\eta_i$, mining of new blocks is described by a Poisson process, i.e. node $i$ has a probability $\mathcal{P}(l,\eta_i t)=e^{-\eta_i t}(\eta_i t)^l/l!$ to mine $l$ blocks in a time $t$. The inter-event time distribution is given by the time derivative of the complementary probability to not mine any block $p_i(t) = 1-e^{-\eta_i t}$ up to time $t$.

Therefore, the time it takes  node $i$ to mine a new block is distributed according to an exponential distribution with parameter $\eta_i$. 
At the global level, the difficulty is adjusted to ensure that new blocks are mined at a constant rate. This \textit{mining interval}  is a design choice that is an arbitrary decision during the system design; for example, $\tau = 10'$ for Bitcoin and its hard forks, $\tau = 12''$ for Ethereum. 
Thus, it is immediate to derive that 
$
\sum_i \eta_i = {\tau}^{-1}.
$
The distribution has the direct implication that the number of blocks mined by node $i$ is distributed according to a Poisson distribution with parameter $\eta_i$. Because of the well-known properties of this distribution, the number of blocks mined by all network nodes are distributed according to a Poisson distribution with parameter $\sum_i \eta_i$.

\paragraph{Blockchain mechanics.} Let $b$ be a \textit{block} (i.e. a bundle of verified transactions). Each new block to be generated must contain information about its parent one (the previous block). To account for this link, $b$ is endowed with a height $\mathtt{h}_b$ (i.e. the number of previous blocks starting from the first one: the \textit{genesis} block), and a discovery time $\mathtt{t}^0_b$ by a miner $\mathtt{m}_b$. 
At time $t$, each node $i$ holds a local replica of the ledger,  meaning that they have created or accepted a set of blocks, let us say $\mathcal{B}_i(t)$. The height of $i$'s blockchain is: 
$
h_i(t) = \max_{b \in \mathcal{B}_i(t)} \left(  \mathtt{h}_b \right).
$

After being accepted by a node, a block is broadcast through the logical peers network. As previously described, the \textit{longest chain rule} specifies that if a block $b$ is broadcast 
from  node $j$ to $i$ 
it gets accepted (added to the local copy of $i$) iff $h_i(t) < h_j(t)$. 
All preceding blocks  are also  accepted, effectively turning the local replica of the $i$'s ledger equal to $j$'s one. The time at which block $b$ becomes accepted by node $i$ is denoted by $\mathtt{t}_b(i)$. When the local replicas of the ledgers holds by the network nodes are all the same, it is said that the system has reached consensus.


The exact time it takes for a block to be broadcast and reach all the nodes in the network depends on multiple factors: the actual distance in the infrastructure network, network traffic, other tasks that may be consuming -- or diverting -- resources away from this communication, the number of transactions included in the block (which affects the block's size), etc. Encompassing all these factors goes beyond the scope of this work. Therefore, out of parsimony and wlg, we consider that the number of blocks relayed from one node to its neighbours follows a Poisson distribution with parameter $1/\tau_{nd}$. Thus, the expected time for $i$ to gossip a block to one of its neighbours is exponentially distributed.
Realistic values for the P2P networks' latencies $\tau_{nd}$ can be found in \cite{gencer2018decentralization}. 
In the numerical simulations, we use $\tau_{nd}$ as a control parameter for the emergence of consensus.

\paragraph{Model simulation.} 
Because of the nature of the proposed parameters, our equations can be efficiently modelled in terms of the Gillespie algorithm \cite{GILLESPIE1976403,gillespie1977exact}. At time $t$, each node $i$ contains a local replica of the blockchain ledger $\mathcal{B}_i(t)$, commonly initialised at time $t=0$ with a single, genesis block. The events that trigger changes at the node level in the system are
\begin{enumerate}
    \item \textit{Block creation}: each node $i$ produces new blocks at a constant rate $\eta_i$. In absence of malicious agents, all nodes are independent Poisson processes, which implies that block creation by \textit{any} node occurs at an aggregate rate $\sum_i \eta_i$. 
    \item \textit{Block diffusion}: at a rate $\tau_{nd}^{-1}$, node $i$ transmits its local replica of the blockchain ledger to node $j$. As per the longest chain rule, if $h_i(t) \leq h_j(t)$ nothing happens. So, without affecting the results, block diffusion can only take place over \textit{directed active links} $i\to j$ where $h_i(t) > h_j(t)$. For the sake of simplicity, we assume that $\tau_{nd}^{-1}$ is the same for all edges. This implies that a diffusion event between \textit{any} edge occurs at an aggregate rate  $E_{a} \tau_{nd}^{-1}$, where $E_{a}$ is the number of active edges. 
\end{enumerate}
Being all events independent, the rate at which \textit{any} event occurs, i.e. the total rate of transitions, is  $\xi =  \sum_i \eta_i + E_{a} \tau_{nd}^{-1}$. To simulate the system, we first select the next event that will take place with a probability proportional to the rate at which it occurs. I.e. the next event will be block creation with probability ${\sum_i \eta_i }/\xi$, 
or block diffusion otherwise. If the next event is ``block creation'', \textit{the node} that creates the block will be selected at random from a probability distribution such as $\eta_i / \sum_j \eta_j$. If the next event is ``block diffusion'', \textit{the link} through which block diffusion starts will be selected with equal probability (because $\tau_{nd}^{-1}$ is the same for all links). Then, time is updated to $t + t'$  where $t'$ follows and exponential distribution with parameter $\xi$ (i.e. the time of the next event is drawn at random from the aggregate rate for all events).

It is important to realise that the results are independent of the exact value of the actual parameters, but they depend only on the relative ratio $ \tau_{nd} / \tau $. This is the ratio between the network delay and the expected time between block discoveries.

\paragraph{Emergent properties}

Our main goal for the emergent properties analysis is to study the emergence of blocktrees (i.e., forks): the full set of blocks that are mined concurrently by the different nodes in the network. The topology of this blocktree serves as a performance indicator of the system's architecture. To detect the transition from consensus to congestion we compare our results with the observed data from different blockchain-based systems.

From the set of all blocks created (denoted by $\mathtt{B}$), a subset of them (denoted by $\mathtt{M}$) is included in the local replica of all the nodes in the network. $\mathtt{M}$ is the set of blocks on which the entire network achieved consensus. All other blocks (denoted by $\mathtt{O}$) are termed off-chain (or orphaned) blocks. At a global level, this allows us to compute the ratio of main-chain blocks $|\mathtt{M}| / |\mathtt{B}|$ and the the ratio of off-chain (or \textit{orphaned}) blocks as
\begin{equation}
\Xi = {|\mathtt{O}|}/{|\mathtt{B}|}
  \label{orphanedblockrate}
\end{equation}
For a set of system parameters, the consensus will be efficient if $|\mathtt{M}| / |\mathtt{B}| \to 1$. 
A related quantity is the branch rate which measures how often forks occur. This is an baseline indicator that a newly created block will not be appended to the main-chain. The ratio of branches in the system is computed as
$
  F = \frac{1}{\left| \mathtt{M} \right|} \sum_{b \in \mathtt{M}} \sum_{c \in \mathtt{O}} \delta\left(  z_b, z_c \right)
$,
where $z_b$ represents the parent of block $b$ and $\delta(\cdot, \cdot)$ is the Kronecker delta. 
Additional properties that quantify the system efficiency and measure the blocktree level of branching are the mean branch length $L$ and the time to create blocks that are attached to the main-chain $T$.


The probability to find the system in global consensus can be defined as the ratio between the time all the nodes have the same local replica of the blockchain $T_c$ and the total simulation time
\begin{equation}
 P = T_c /T_{sim}.
\label{probconsensus}
\end{equation}
We use this measure in the numerical simulations as the order parameter for the transition to consensus.


To measure the power concentration of block creation we use the Gini index. By denoting the number of blocks created by node $i$ as $w^{bt}_i = \sum_{b \in \mathtt{B}} \delta(i, \mathtt{m}_b) $, the Gini index of the distribution of (sorted) values $G(w^{bt}) = {2 \sum_{i}^n iw^{bt}_i}/{n \sum_{i}^n w^{bt}_i} - (n+1)/{n}$, is an indicator of power concentration of blocks. For the latter we can distinguish between blocks that end up in the main chain $w^{mc}_i = \sum_{b \in \mathtt{M}} \delta(i, \mathtt{m}_b) $ or off-chain $w^{oc}_i = \sum_{b \in \mathtt{O}} \delta(i, \mathtt{m}_b) $ resulting in $G_{mc}=G(w^{mc})$ and $G_{oc}=G(w^{oc})$, respectively.
Analogously, we can estimate the level of heterogeneity of different hashing/staking power distributions with the corresponding index $G_\pi=G(\pi)$, where $\pi_i$ is the computational/staking power of node $i$.



\section{Results}

\subsection{Data analysis and model selection}
\label{sec:data_analysis}

In order to fine-tune our model we analyse the mining data for the following three cryptocurrrencies: Bitcoin, Ethereum and Litecoin, on a yearly basis ranging from year 2013 to 2019.
For all platforms, we analyse the revenue share per miners based on the fraction of blocks that each miner created over the period 2013--2019. Since we observe a large volatility of the revenue share per block, instead of determining the probability  $\mathcal{P}(\pi)$ that an arbitrary miner has computational hashing (or staking) power $x$ -- in an infinitesimally small interval around a given value $\pi$ --, we compute the survival function or complementary cumulative distribution function (CCDF) $P(x>\pi) = \int_{\pi}^{\infty}\mathcal{P}(\pi')d\pi'$.
Then for power-law distributed hashing (staking) powers $\mathcal{P}(\pi)\sim \pi^{-\alpha}$, with $\alpha\ne1$, we also obtain a power-law for the CCDF but with exponent $\beta=1-\alpha$.
However, since the maximum value of the hashing/staking power $\pi_{max}$ is considerably small even for power-law distributed hashing/staking powers, we obtain significant deviations from a power-law in the CCDF with strong cutoffs.

\begin{table}[h]
\centering
\caption{\label{tab:table1_hp}Model selection for the distributions of the miners' annual hashing power for the three cryptocurrencies analysed. 
The columns are: Log-likelihood ratio $R$ and the corresponding $p$-value for power-law  and exponential distributions. The power-law exponent $\alpha$ is obtained from the numerical fit of the real data using a power-law distribution $\mathcal{P}(\pi)\sim \pi^{-\alpha}$ and the average $\lambda=\braket{\pi}$ of the  exponential distribution $\mathcal{P}(\pi)= \lambda^{-1} \exp(-\pi/\lambda)$. Highlighted in bold are the significant values of the log-likelihood test.
}
\begin{tabular}{ccccccccccccc}
 &\multicolumn{4}{c}{Bitcoin}&\multicolumn{4}{c}{Ethereum}&\multicolumn{4}{c}{Litecoin}\\ \hline
 Year&$R$&$p$&$\alpha$&$\lambda$&$R$&$p$&$\alpha$&$\lambda$&$R$&$p$&$\alpha$&$\lambda$\\ \hline
 2013 & -2.9 & 0.18 & 2.75 & 0.06 &  &  &  &  & -0.36 & 0.0 & 12.88 & \bf{0.57}\\
 2014 & -11.57 & 0.05 & 1.84 & 0.05 &  &  &  &  & -0.18 & 0.16 & 14.4 & 0.28\\
 2015 & -15.74 & 0.0 & 2.09 & \bf{0.04} & 1328.65 & 0.0 & \bf{1.68} & 0.0 & -0.55 & 0.03 & 7.35 & \bf{0.31} \\
 2016 & -2.98 & 0.01 & 3.58 & \bf{0.04} & 5050.29 & 0.0 & \bf{1.43} & 0.0 & -0.02 & 0.8 & 28.17 & 0.23 \\
 2017 & -5.99 & 0.0 & 3.14 & \bf{0.03} & 3048.54 & 0.0 & \bf{1.29} & 0.01 & -7.66 & 0.07 & 1.92 & 0.11 \\
 2018 & 0.79 & 0.57 & 4.46 & 0.05 & 192.34 & 0.0 & \bf{1.57} & 0.01 & -0.2 & 0.57 & 9.57 & 0.11 \\
 2019 & -0.62 & 0.34 & 5.79 & 0.06 & 131.12 & 0.0 & \bf{1.67} & 0.01 & 5.4 & 0.02 & \bf{3.27} & 0.1 
\end{tabular}
\end{table}

For the model selection of the the miner's annual hashing power, we have compared three different distributions: 1) pure power-law with no cutoffs; 2) power-law with exponential cutoffs and; 3) exponential distributions. While the power-law distribution with exponential cutoffs provides a better data description than the pure power-law, we will consider only the latter and the exponential distribution since they are equally comparable with respect to the observed data.  
In Table~\ref{tab:table1_hp} we show the results of the model selection analysis of the distributions of annual hashing powers for the three cryptocurrencies in our sample.
In each column, the log-likelihood ratio $R$ is reported, as well as the corresponding $p$-value for the pair of  distributions used to fit to the revenue share data, using a power-law and an exponential distribution. A value of $R$ larger than 0, implies that the power-law distribution is preferred to the exponential distribution and vice versa if $R$ is negative. If $p<0.05$ the sign of $R$ is taken to be significant; otherwise, it is taken to be the result of statistical fluctuations.

For Bitcoin, we find that the exponential distribution is a substantially better fit for the data with respect to a power-law across all spanned time periods, with the exception of year 2018, but with a very large $p-$value making the value of R non-significant.
Contrary, for Ethereum, we find that a power-law is consistently a better fit than an exponential. For Litecoin, we find mixed results when the values of $R$ can be taken as significant.

The distribution parameters estimation from the numerical fit of a power-law and of the mean of the exponential distribution are reported in the same Table. We find that, for Ethereum, where the probability distribution function of the data is well-fitted by a power-law, an appropriate choice for the exponent is $\alpha \approx 1.5$.
Instead, from the revenue shares of Bitcoin we find that the data is better distributed as an exponential with average $\lambda = \braket{\pi} \approx 0.05$.

\begin{figure}[]
\centering
\includegraphics[width=0.3\textwidth]{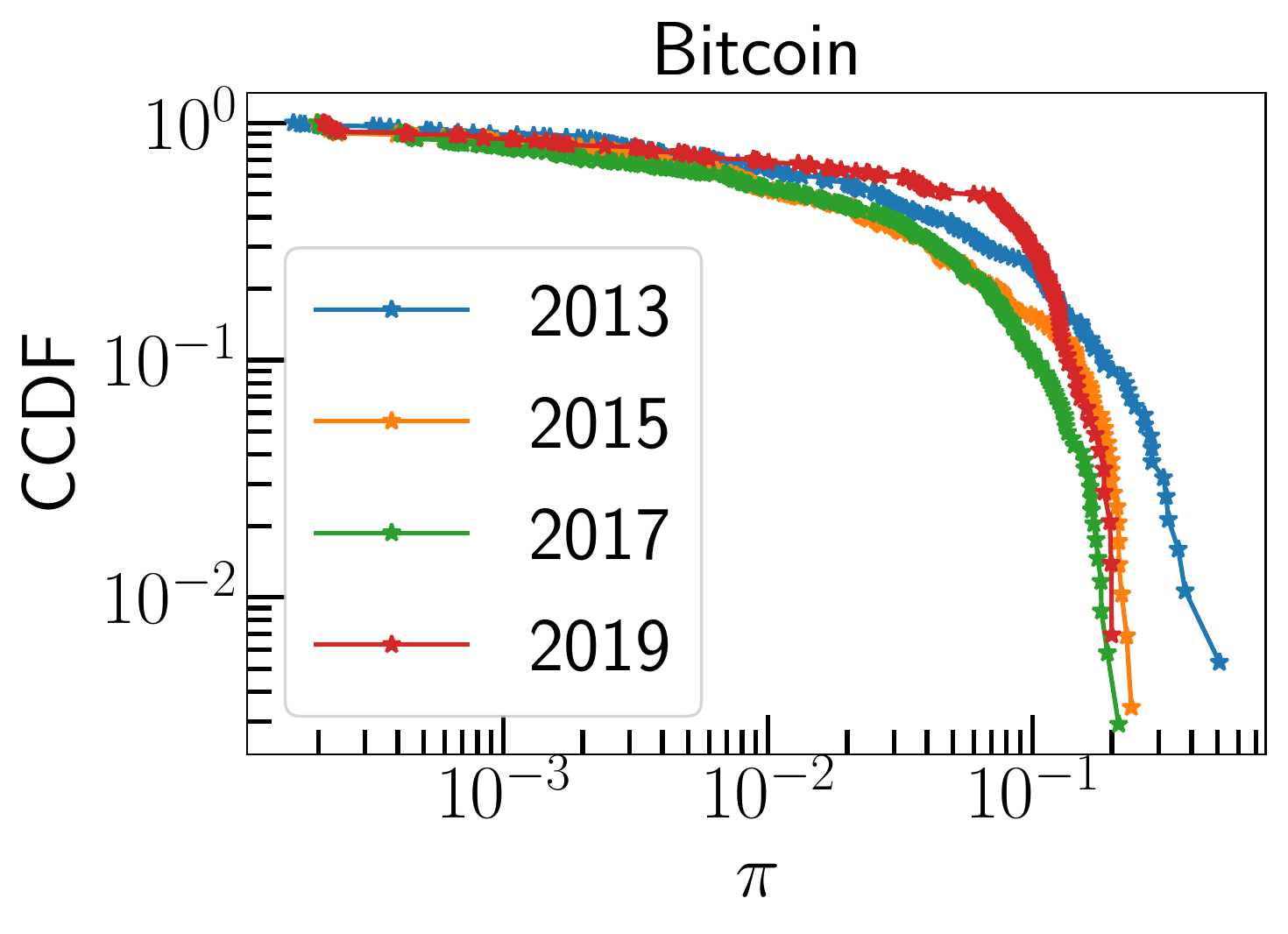}
\includegraphics[width=0.3\textwidth]{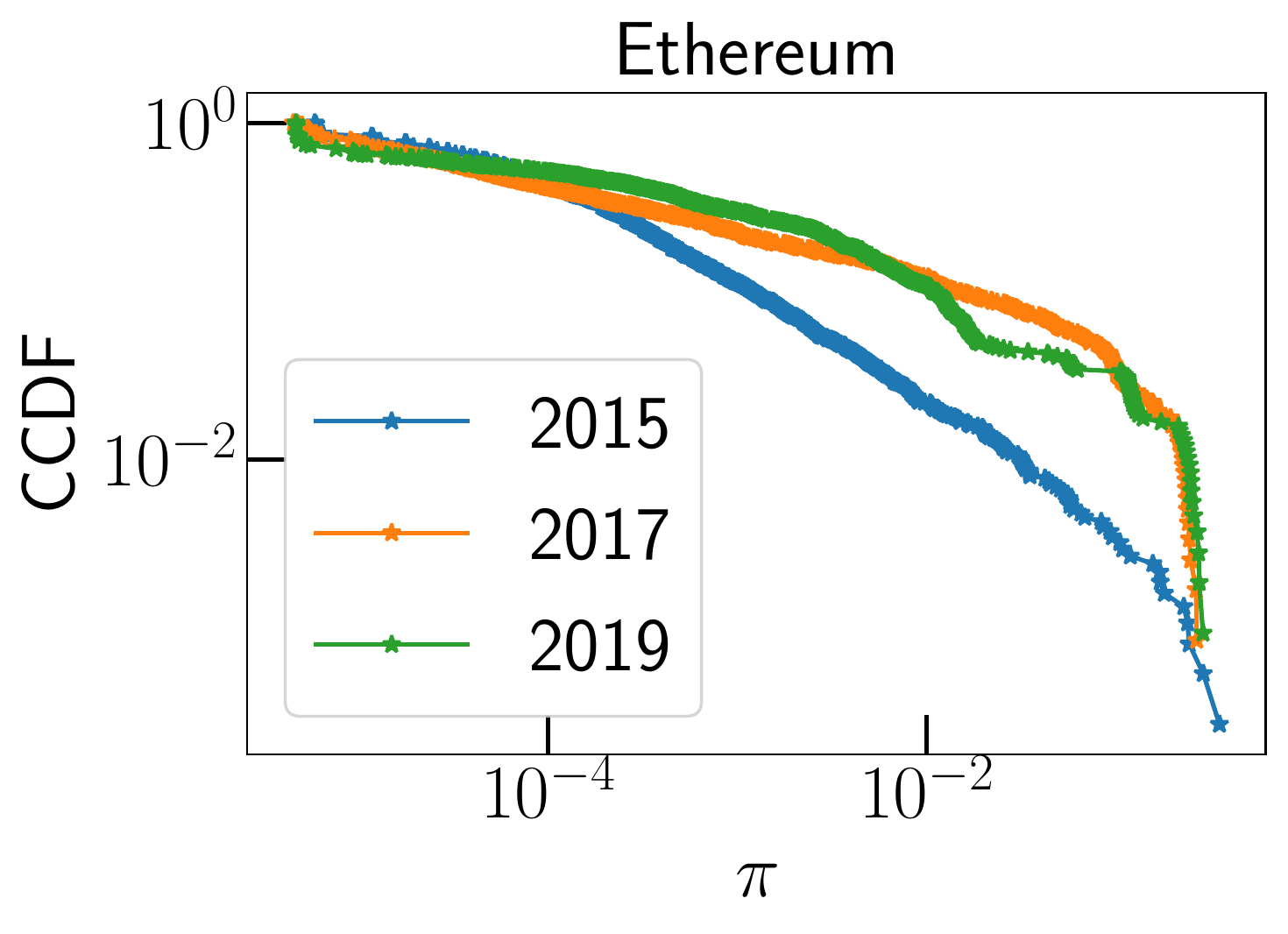}
 \includegraphics[width=0.3\textwidth]{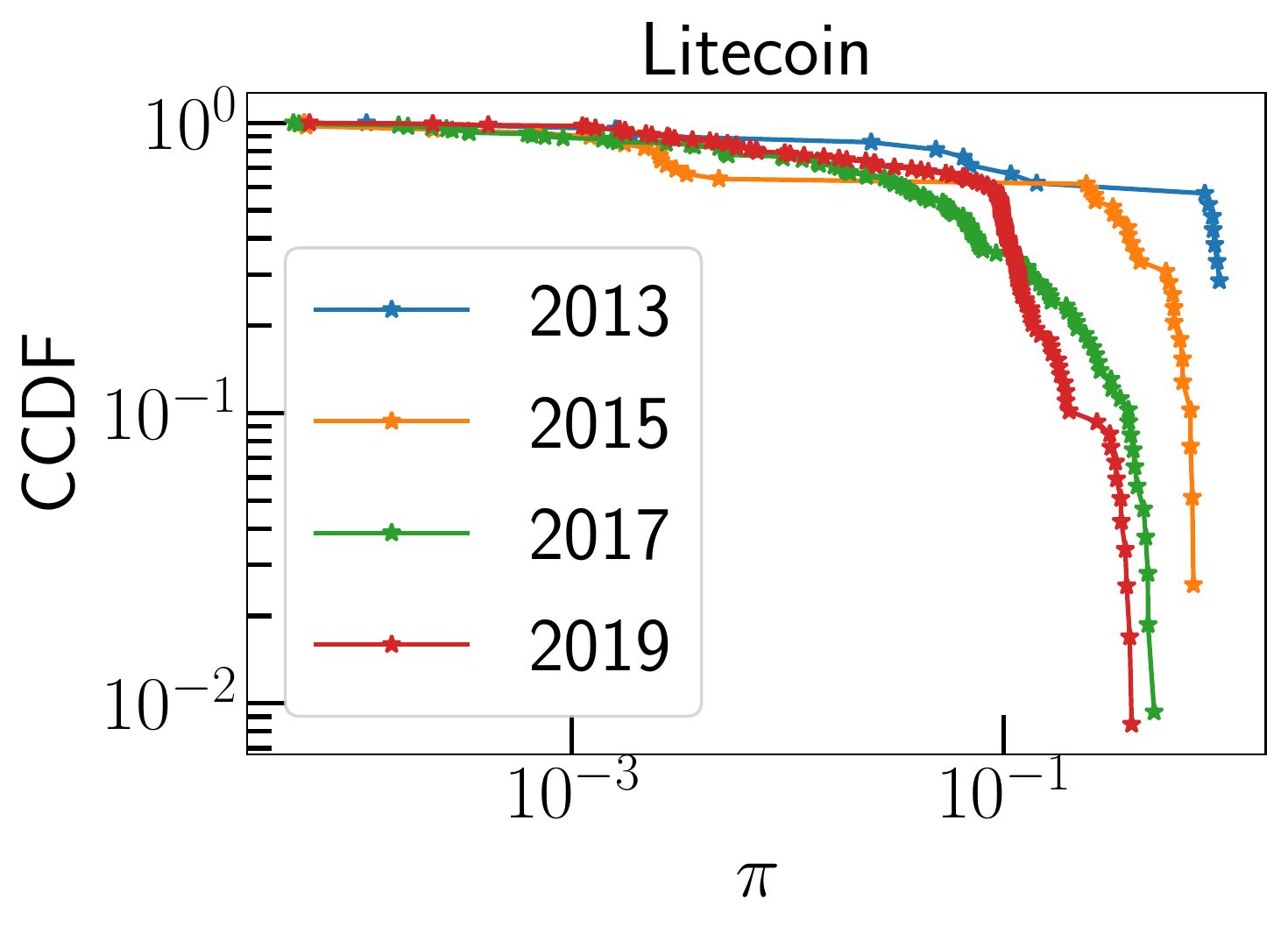}
    \caption{\label{fig:data}
    Yearly complementary cumulative distribution function of the computing hashing power $\pi$ for Ethereum, Bitcoin and Litecoin. 
   }

    
\end{figure}

In Fig.~\ref{fig:data} we show the complementary cumulative distribution function of the annual shared revenue of blocks $\pi$ for Ethereum from 2015 to 2019, and for Bitcoin and Litecoin in the period from 2013 to 2019. The data displays wide heterogeneity, that spans across many orders of magnitude, with significant cutoffs for high values of $\pi$.

\subsection{Analytical estimation of the branching threshold}
\label{sec:analytical}

In this subsection we are interested in modelling the relationship between the mining process and the diffusion process. 
Starting with the block diffusion process, we can use either the direct-path propagation model, or the multiple-path propagation model to estimate the arrival times for the diffusion of blocks in the P2P network.

For the direct-path propagation model, we need to assume that loops in the P2P network are irrelevant and that the graph is locally tree-like. In this case, the distance between nodes is based on the single shortest path and it can be used as a proxy for the mined block diffusion times from a source node $i$ to a target node $j$ during the mining process. To measure the block arrival time $T_{ij}$, we can use a heuristic to estimate $T_{ij}$ by using the number of links between $i$ and $j$ given by the shortest path distance $D_{ij}$ weighted by the block diffusion delay $\tau_{nd}$, such that: $T_{ij} = \tau_{nd} D_{ij}$. 
The average $\langle T_{ij} \rangle = \tau_{nd} \langle D \rangle$, where $\langle D \rangle$ is the average shortest path in the network, gives a heuristic mean-field estimation of the threshold delay defining a sub-optimal branched state.

The previous approach can be improved by considering the multiple-path propagation model that accounts for the loops which emerge locally, and the multiplicity of paths connecting the miner of a block with the other miners. However, to find all simple paths connecting two nodes in the graph is a non polynomial problem and becomes computationally unfeasible for large graphs. 
To overcome this issue, we can relax the assumption of direct path propagation and leverage on random-walk theory to estimate the minimum diffusion distance between two nodes. The quantity that estimates the random-walk distance in a network is the mean-first passage time $M_{ij}$ between node pairs \cite{kemeny1960finite}.
Given the adjacency matrix of the P2P network $A_{ij}$, the probability that a block is passed from node $i$ to node $j$ in one time step is given by the Markov transition probability matrix $P_{ij}=A_{ij}/k_i$, where $k_i=\sum_j A_{ij}$ is the degree of node $i$.
If nodes $i$ and $j$ are connected, the block is immediately passed in the next time step; thus, the diffusion time is equal to $1$. Otherwise, the time for a block to reach $j$ is given by the average diffusion time from all other nodes $k$, excluding the target $j$, weighted by the corresponding diffusion probability. The total contribution is given by the mean-first passage time $M_{ij}=1+\sum_{k\ne j} P_{ik}M_{kj}$.  
In this scenario, $T_{ij} =\tau_{nd} M_{ij}$ and by averaging over all nodes we obtain $\langle T_{ij} \rangle = \tau_{nd} \langle M \rangle$, where $\langle M \rangle = \sum_{ij} M_{ij}/(N(N-1)$ is the average of the global mean-first passage time.
By using the mean-first passage time approach we just discussed, we provide an estimation of the branching diffusion time $\tau_b$ on the P2P network, such that the maximum mining time is below the threshold given by the inter-block times $\tau$.
In the following, we compare the numerical simulation of the consensus protocol with the estimation of the branching threshold given by  
\begin{equation}
\tau_b = {\tau}/{\langle M \rangle}. 
\label{taub}
\end{equation}
The transition between optimal consensus with blocks belonging all in the main chain and the branched state is described by the orphaned block rate $\Xi(\tau_{nd})$ parameter defined by Eq.~\eqref{orphanedblockrate}, as a function of the block diffusion delay $\tau_{nd}$.
In the next Section we show that, for two characteristic network topologies (Erdos-Renyi and Barabasi-Albert),  $\Xi(\tau_{nd})$ vanishes for values of $\tau_{nd}$ that are below the branching threshold $\tau_b$.


\subsection{Numerical results}

In this Section, we present the results of the numerical simulations of the blockchain consensus protocol using the model described in Section \ref{sec:modelling}.
We fix the number of simulation steps to $T_{sim} = 20,000$ and average all quantities over many independent stochastic realisations of the process of block creation and diffusion.

In addition, we analyse two emergent properties of the system in detail. First, we study the emergence of branches, the onset of which can be estimated analytically from the sole knowledge of the network topology of the miners, as described in Section \ref{sec:analytical}.
We compare the analytical estimation of the branching threshold $\tau_b$, given by the reciprocal average random-walk hitting times, when the fraction of orphaned blocks becomes non-negligible.
Second, we identify a phase transition characterising the consensus process. It is described in terms of the order parameter $P(\tau_{nd})$, which quantifies the probability for the all the nodes (miners) in the system to have the same local replica of the blockchain, when the network latency (control parameter) $\tau_{nd}$ assumes different values,  see Eq.~\eqref{probconsensus}. By performing a finite size scaling analysis, we numerically compute the critical diffusion delay $\tau_c$, above which consensus is not possible.

In Section \ref{sec:analytical} we modelled the block diffusion propagation path in order to derive the branching threshold. Here, we extend our study by analysing the fraction of orphaned blocks ($\Xi$), the fraction of branches with respect to the main chain ($F$) and the probability of global network consensus ($P$) with respect to two major network topologies, namely, the Erdos-Renyi (ER) and Barabasi-Albert (BA) topologies.

Let us see why we choose the ER and BA networks. In real cryptocurrency networks running the respective P2P protocol, mining nodes act in competition to solve the PoW algorithm.  Miners can also cluster together, forming so-called mining pools or farms, which increases their probability of success. The dynamics of the P2P network's formation can be summarised as follows. New nodes enter the system, i.e. the current version of the P2P network, and generate links with randomly chosen existing miners, thus making the network to mimic a random network. 

In this scheme, old nodes have the advantage of being chosen more often than new nodes introducing ageing effects. This mechanism effectively introduces a deviation from a homogeneous to a more heterogeneous topology, which can be modelled using scale-free networks \cite{caldarelli2007scale}.
Although the exact real topology of the mining network is not known by design for any cryptocurrency, relevant network properties have been extensively studied \cite{miller2015discovering,javarone2018bitcoin,delgado2019txprobe}.
In particular in \cite{javarone2018bitcoin}, for Bitcoin and Bitcoin Cash networks, the authors found evidence of both scale-free and small-world properties. We do not have solid reasons to advance the hypothesis that other blockchains' mining networks deviate substantially from these in terms of network topology.
Thus, we simulate the consensus protocol using two relevant synthetic graph topologies as representative of  homogeneous and heterogeneous degree distributions, namely the ER and BA topologies. We then use these models as benchmark for the unknown P2P random network topologies.

Without loss of generality, in the following we set the mining interval to be $\tau=1$. 
For the BA topology, we study dependence on the size of the network. For the ER topology, we consider instead how the average connectivity in the graph $\braket{k}$, or equivalently the probability for random edge creation between nodes, affects both the block diffusion process and the transition to consensus. 
Motivated by the values obtained from the fit of the real mining data (see Section \ref{sec:data_analysis}), we consider - for both BA and ER - the hashing power of the miners to be either a power-law $\mathcal{P}(\pi)\sim\pi^{-\alpha}$ with exponent fixed to $\alpha=1.5$, or an exponential distribution $\mathcal{P}(\pi)= \lambda^{-1} \exp(-\pi/\lambda)$ with mean $\lambda=\braket{\pi}=0.05$. 

\begin{figure}[h!]
\begin{center}
\includegraphics[width=0.3\textwidth]{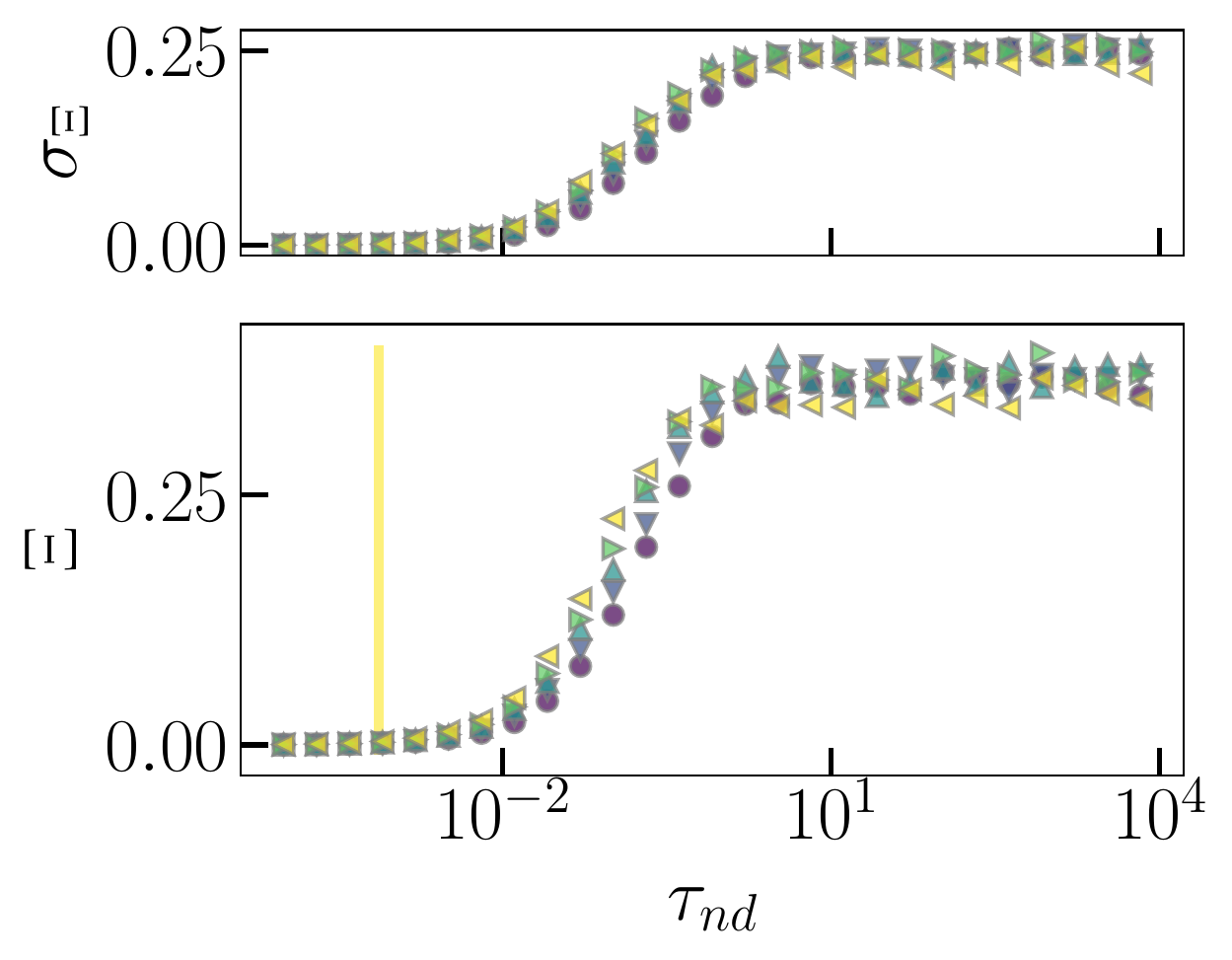}
\includegraphics[width=0.3\textwidth]{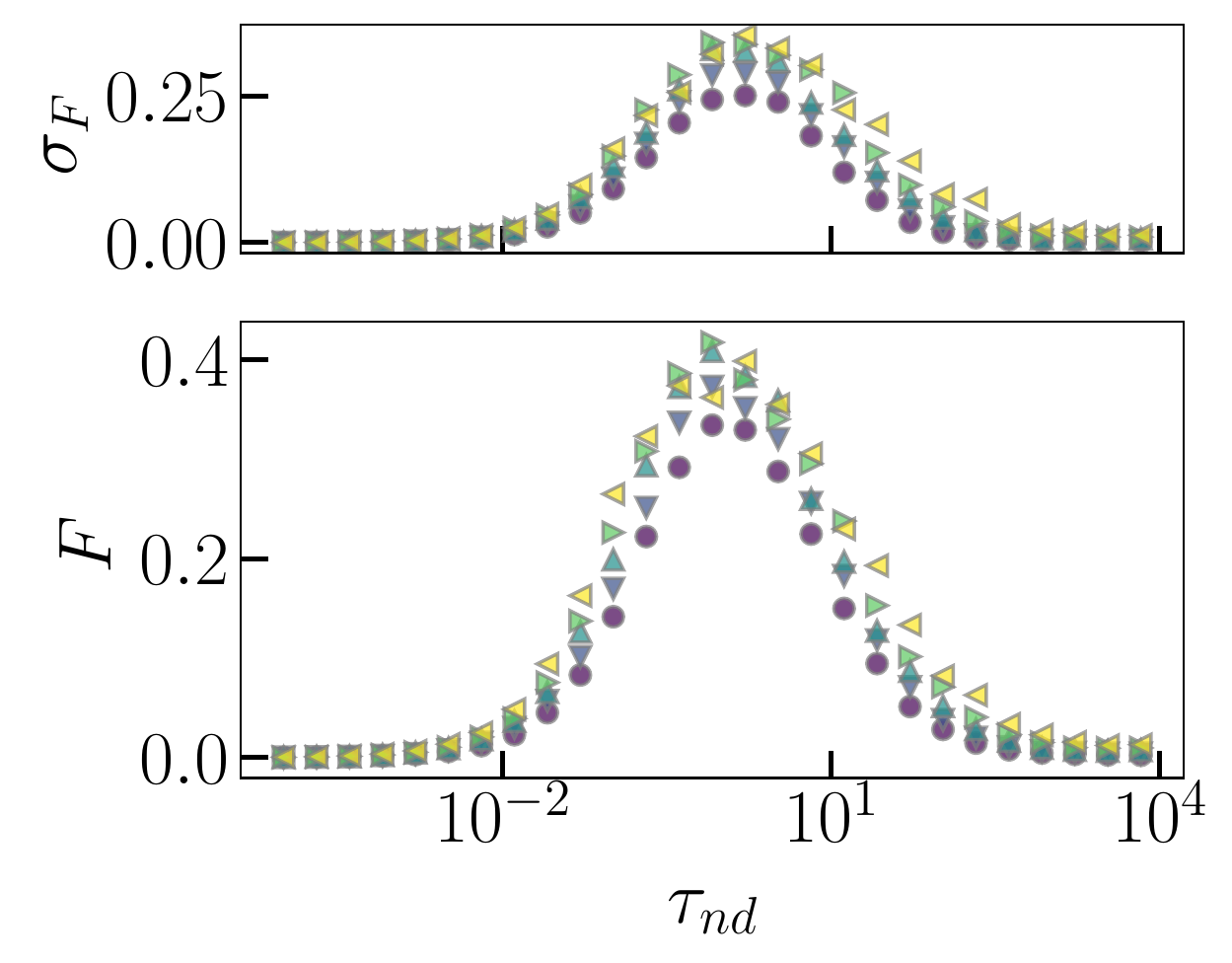}
\includegraphics[width=0.3\textwidth]{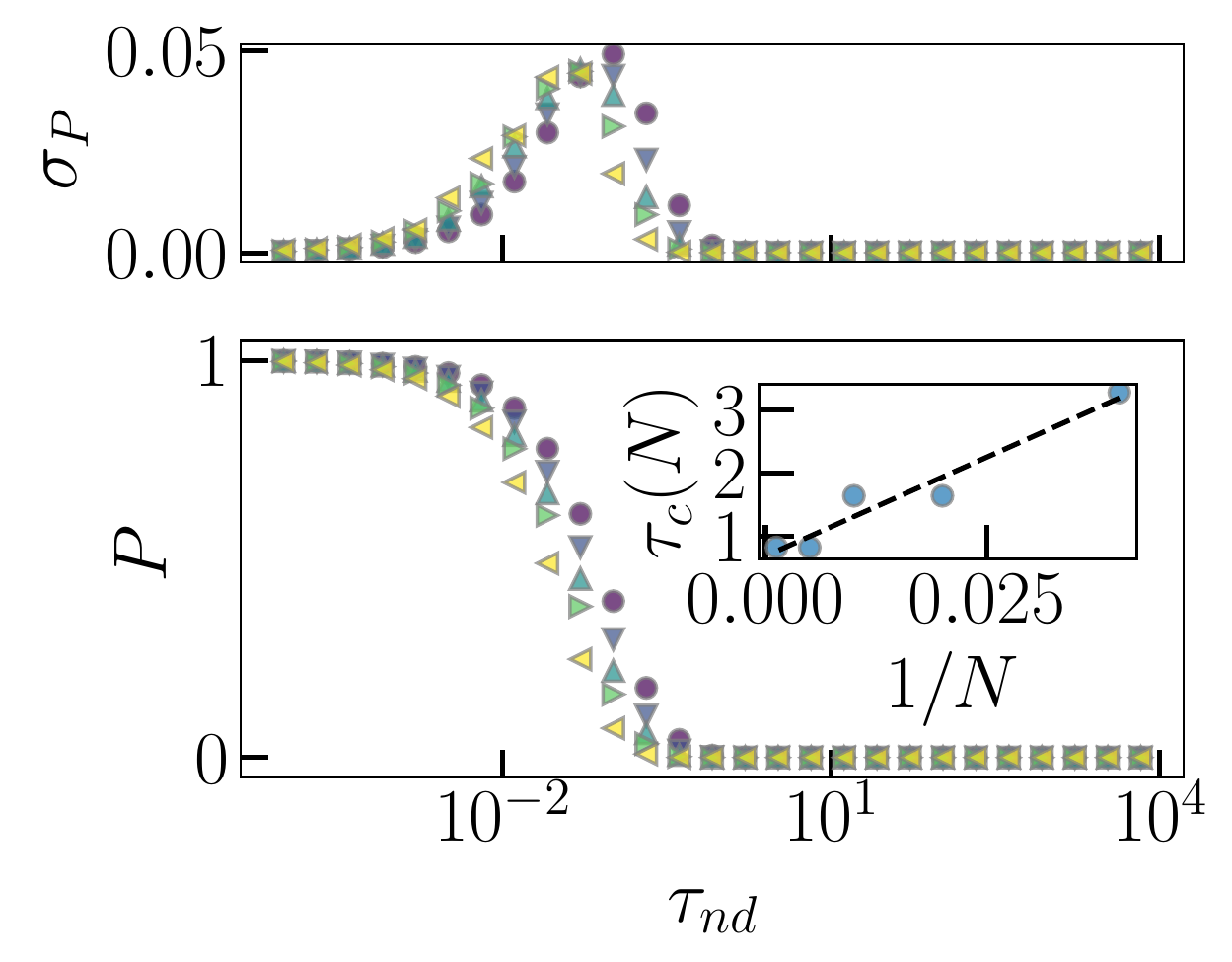}

\includegraphics[width=0.3\textwidth]{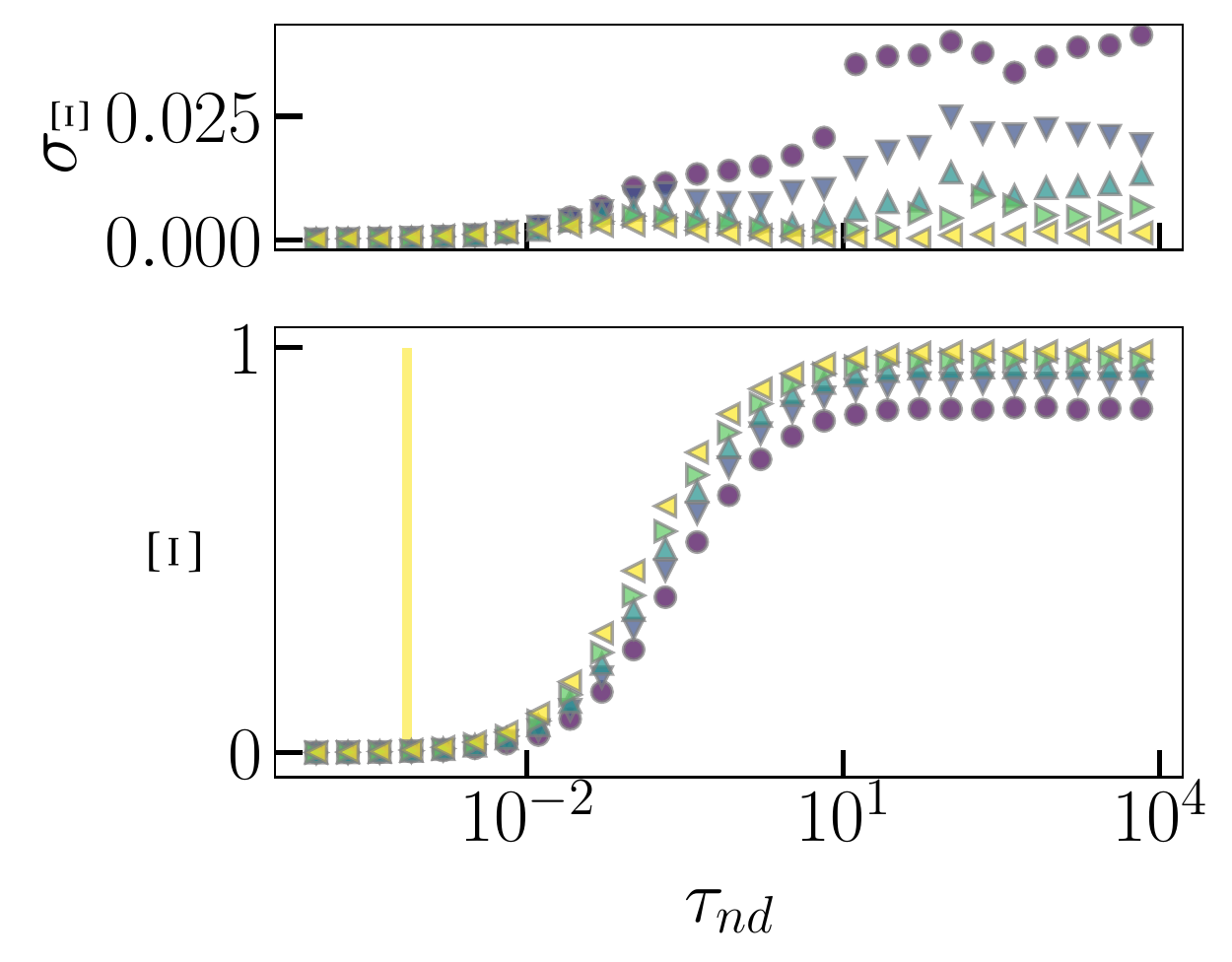}
\includegraphics[width=0.3\textwidth]{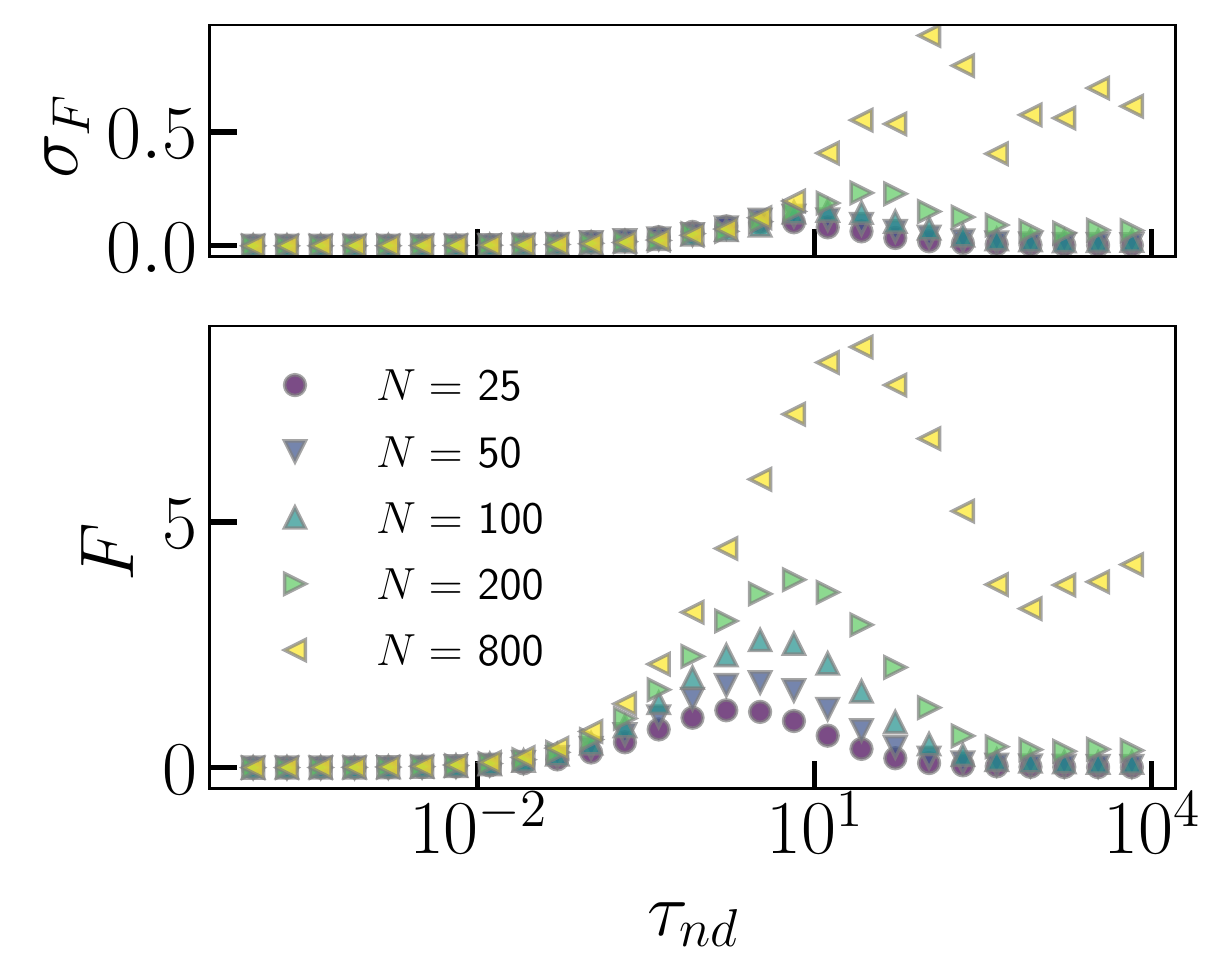}
\includegraphics[width=0.3\textwidth]{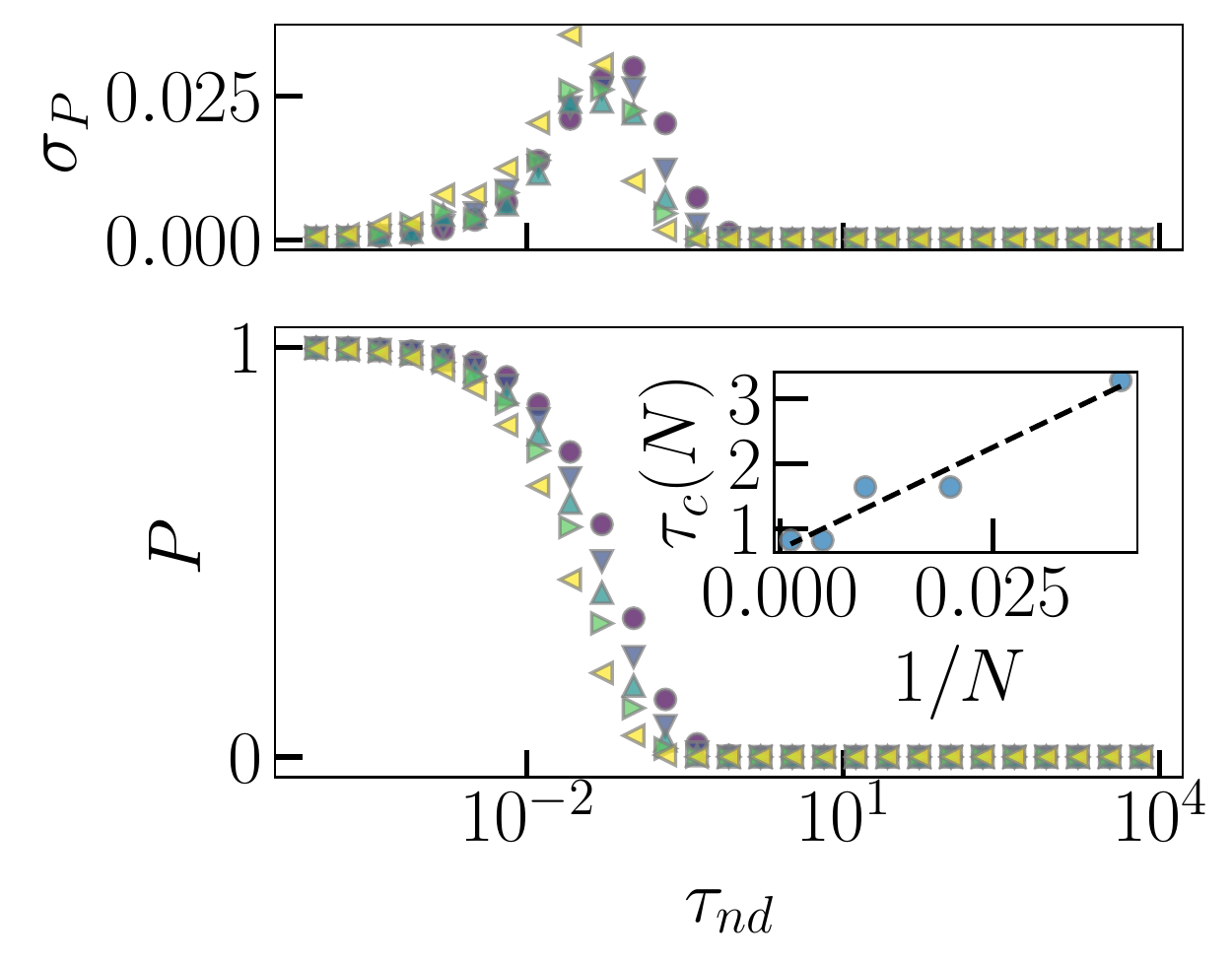}
\end{center}
\caption{\label{fig:3}
Results for the power-law (top) and exponential (bottom) hashing power distributions, with realistic parameters $\alpha=1.5$ and $\lambda=0.05$ for P2P networks with BA topology as a function of the block delay $\tau_{nd}$. 
Different quantities and the respective standard deviations $\sigma$ are: the fraction of orphaned blocks $\Xi$, the fraction of branches with respect to the main chain $F$, and the probability for consensus $P$.
Different colours represent different numbers of nodes $N$ with fixed $m=3$ edges attached from a new node to existing nodes.
The vertical lines in the orphaned block rate panel mark the characteristic branching delay $\tau_b$, defined by Eq.~\eqref{taub}.
Insets of the right panels: finite size scaling of the consensus critical delay, yielding $\tau_c \approx 0.68$.}
\end{figure}

\begin{flushleft}
\textbf{BA Topology Analysis}. 
Fig.~\ref{fig:3} shows the results of the numerical analysis. From left to right: the fraction of orphaned blocks $\Xi$, the branch rate with respect to $F$, and the probability to reach global consensus $P$, all as function of the block diffusion time $\tau_{nd}$. These dependencies are measured with respect to the BA topology with either power-law or exponentially distributed hashing power (top and bottom panels, correspondingly). Fig.~\ref{fig:3} shows also their corresponding standard deviations (upper panels). All quantities are defined in Section \ref{sec:modelling}.

\begin{itemize}[leftmargin=2cm]
 \item[$\Xi = f(\tau_{nd}$)] By observing the dependency of the orphaned block rate $\Xi$ with respect to the network latency, namely the block diffusion time $\tau_{nd}$ we can identify two regimes that emerge independently from the hashing power of the miners. Depending on the mining network size, we find that for $\tau_{nd} \lesssim \tau_b$, the fraction of orphaned blocks is vanishing: $\Xi \to 0$. See Fig.~\ref{fig:3} - left panels. The branching block delay $\tau_b$ defined in Section \ref{sec:analytical} is marked with colour-coded vertical lines. 
 
On the contrary, for  $\tau_{nd} > \tau_b$ the number of orphaned blocks increases and the blockchain branches. This finding holds for both power-law and exponentially distributed hashing powers. The only difference is that the stationary value approaches the maximum $\Xi(\infty)=1$ only for the exponential hashing power distribution, while for the power-law case we find $\Xi(\infty)\approx 0.4$. This is due to the fact that, for heterogeneous hashing powers, a substantial fraction of blocks are mined by the same nodes and, even if the diffusion delay is large, the system behaviour is dominated by those nodes (namely, big miners).

Although the behavior of orphaned block rate $\Xi$ with respect to the BA network latency is robust for both hashing power distributions, there is more uncertainty in the case of power-law distribution than in the exponential one. See Fig.~\ref{fig:3} - left upper panels displaying the standard deviations.

\item[$F = f(\tau_{nd}$)] The fraction of branches with respect to the main chain ($F$) when measured against $\tau_{nd}$ behave similarly than what we have seen for $\Xi$. However, there are significant differences between the power-law and the exponential hashing power distributions. See Fig.~\ref{fig:3} - central panels. For low values of the block delay, we have no branches and $F\approx 0$. At $\tau_{nd} \approx \tau_b$, the blockchain becomes branched with an increasing number of off-chain blocks being validated. 
While the behaviour is qualitatively similar among the two hashing power distributions, $F$ displays a maximum which increases with the number of miners in the P2P network only for exponential hashing power, and is one order of magnitude larger than for the heterogeneous case (power law distribution). See Fig.~\ref{fig:3} - central upper and lower panels. By comparing these results with those obtained for the orphaned block rate (see Fig.~\ref{fig:3} - left panels), we observe that, when $F$ (fraction of branches with respect to the main chain) reaches its peak, $\Xi$ (fraction of orphaned blocks) reaches the level of saturation, which means the blockchain is fully congested with only orphaned blocks.

\item[$P = f(\tau_{nd}$)] For the global consensus probability $P$, we numerically estimate the critical network latency level $\tau_c$, after which consensus is not possible to reach anymore. See Fig.~\ref{fig:3} - right panels. In the insets, we show the finite size scaling analysis to estimate the consensus critical delay $\tau_c$ as a function of the system size. From the fit extrapolation at $N\rightarrow \infty$, we find $\tau_c \approx 0.68$ for both hashing power distributions.

Thus, we identify three regimes:
(i) for $\tau_{nd} \ll  \tau_b < \tau_c$, we have efficient consensus, with a linear blockchain and the system is in consensus for all times, since out-of-consensus time is negligible, and the likelihood of having forks vanishes; 
(ii) for $\tau_{nd} \ge  \tau_c$, the system is never in consensus, with a large fraction of orphaned blocks and 
the branching rate with respect to the main-chain slowly vanishes; 
(iii) for $ \tau_b <\tau_{nd} <  \tau_c$, the system reaches consensus, but is in consensus only for a finite fraction of time, and there is a finite fraction of blocks which has forks.
\end{itemize}
\end{flushleft}

\begin{flushleft}
\textbf{ER Topology Analysis}. For the sake of computational effort we fix the total number of ER nodes to be $N=200$. Here, we analyse the blocktree shape in terms of the average branch length $L$, and its relation to the time needed to add a block to the main chain $T$, both as a function of block diffusion time (network delay) $\tau_{nd}$. For both hashing power distributions (exponential and power/law), Figure \ref{fig:4} shows both quantities ($L$ and $T$) by varying the average degree $\braket{k}$ of the ER network (i.e., the probability to add an edge between two randomly selected miners). In alignment with the previous results, we find that the topology of the blocktree becomes increasingly branched with longer branches. The maximum branch length is found, for both hashing power distributions, when $\braket{k}=1$, with a substantial increase at very large $\tau_{nd}$. This can be explained in terms of the known percolation point of the ER model at exactly $\braket{k}=1$, for which the system starts to fragment in many disconnected components, reducing the possibility of reaching consensus and increasing the branching in the network.

From the corresponding behavior of $T$ we also find that it takes longer time to add blocks to the main-chain and that this is harder for more homogeneously distributed hashing powers (Figure \ref{fig:4} - bottom right panel).
Interestingly, increasing the network connectivity $\braket{k}$ does also increase the time to add blocks to the main chain at fixed block propagation delay when $\tau_{nd}>\tau_c$ (Figure \ref{fig:4} - right panels), i.e. in the out-of-consensus phase, but it also corresponds to significantly lower branch lengths in the blocktree. To get an insight on how the emergent properties are linked to the level of heterogeneity in the system, we compute the Gini index for: (i) the distribution of the miners hashing power $G_\pi$; (ii) blocks added to the main chain $G_{mc}$, and (iii) off-chain blocks $G_{oc}$. 

In Fig.~\ref{fig:5} we show all indices with the respective standard deviations $\sigma$ for power-law (top) and exponential hashing powers (bottom), for the ER topology with $k=8$, and $N=200$ nodes as function of the block delay $\tau_{nd}$. Fig.~\ref{fig:5} (right panels) shows the corresponding fraction of miners per blocks in the blocktree $n_{bt}$, in the main chain $n_{mc}$ and off-chain $n_{oc}$. Different colours represent different average degree $k$ with fixed $N=200$ number of nodes.

For low block delays, the Gini index of blocks attached to the main chain is maximal, as there are no orphaned blocks, and for both heterogeneous and homogeneous hashing powers we have $G_{mc}(\tau_{nd} \ll 1)\approx G_\pi$ and $G_{oc} \approx 0$. 
Increasing the block diffusion delay results in more off-chain blocks and thus increases the value of $G_{oc}$, showing that there is consistent heterogeneity in the mining process. 
In this regime, the system is always not in consensus and fully branched with only off-chain blocks mined by nodes with higher hashing power.

\begin{figure}[]
\begin{center}
\includegraphics[width=0.3\textwidth]{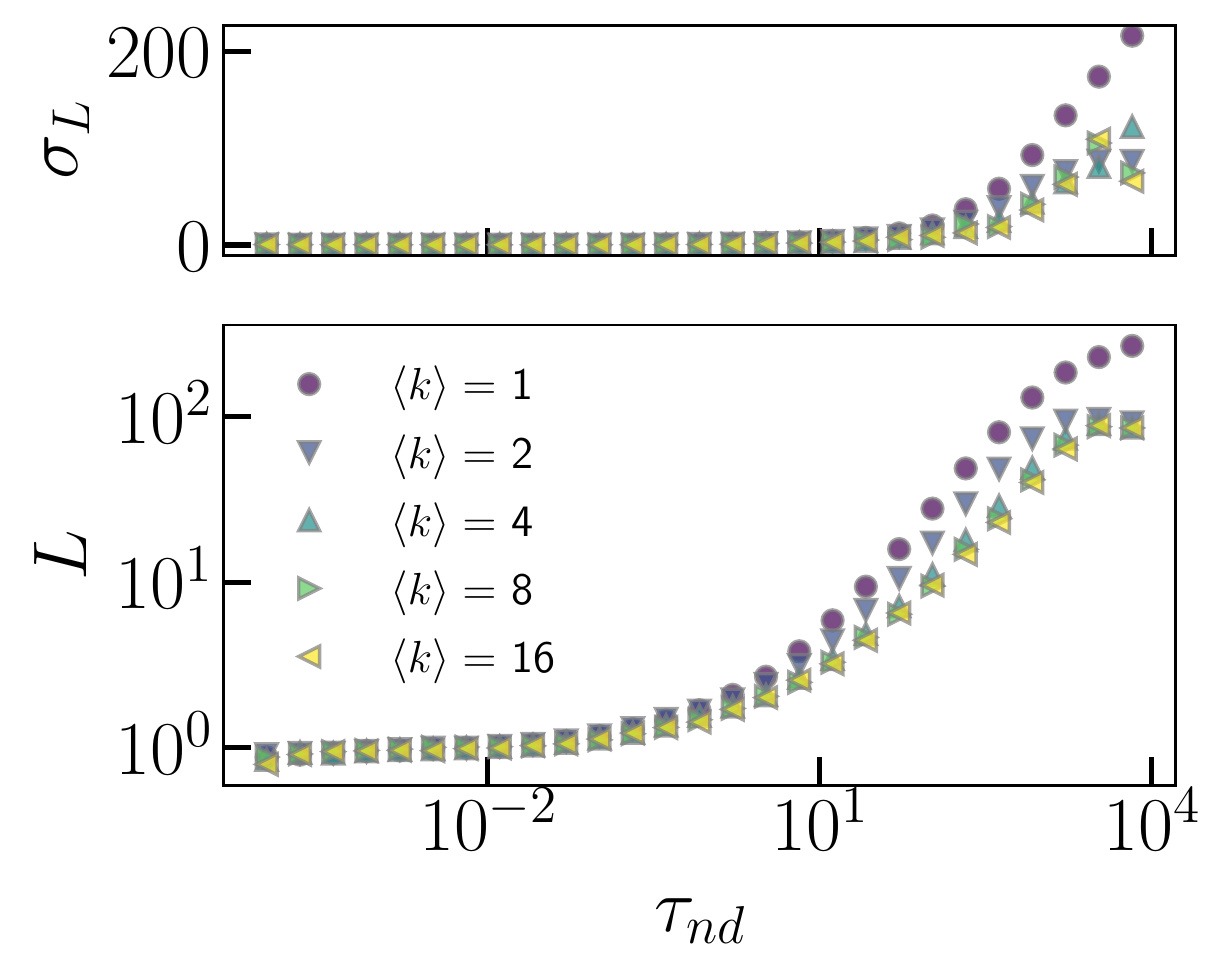}
\includegraphics[width=0.3\textwidth]{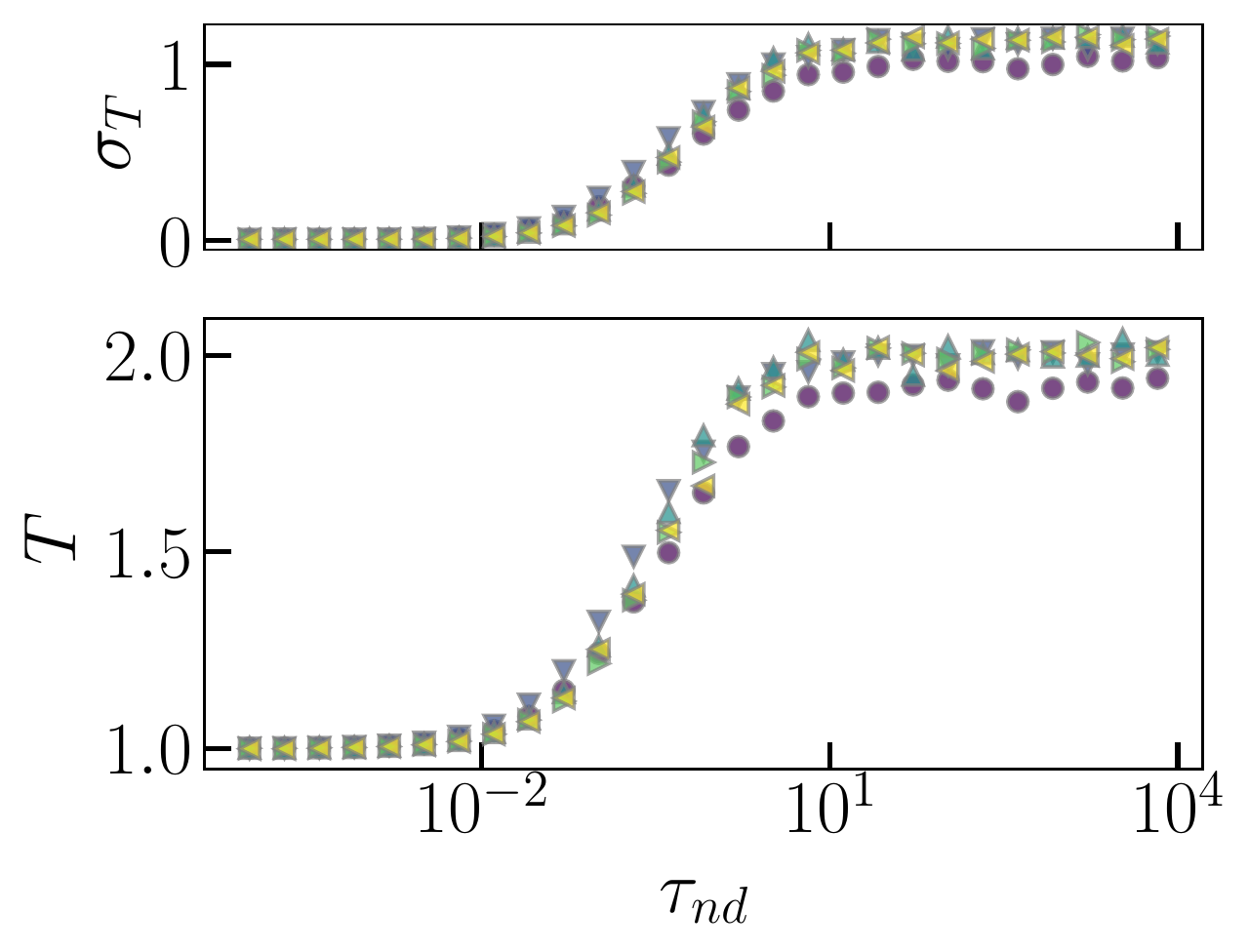}

\includegraphics[width=0.3\textwidth]{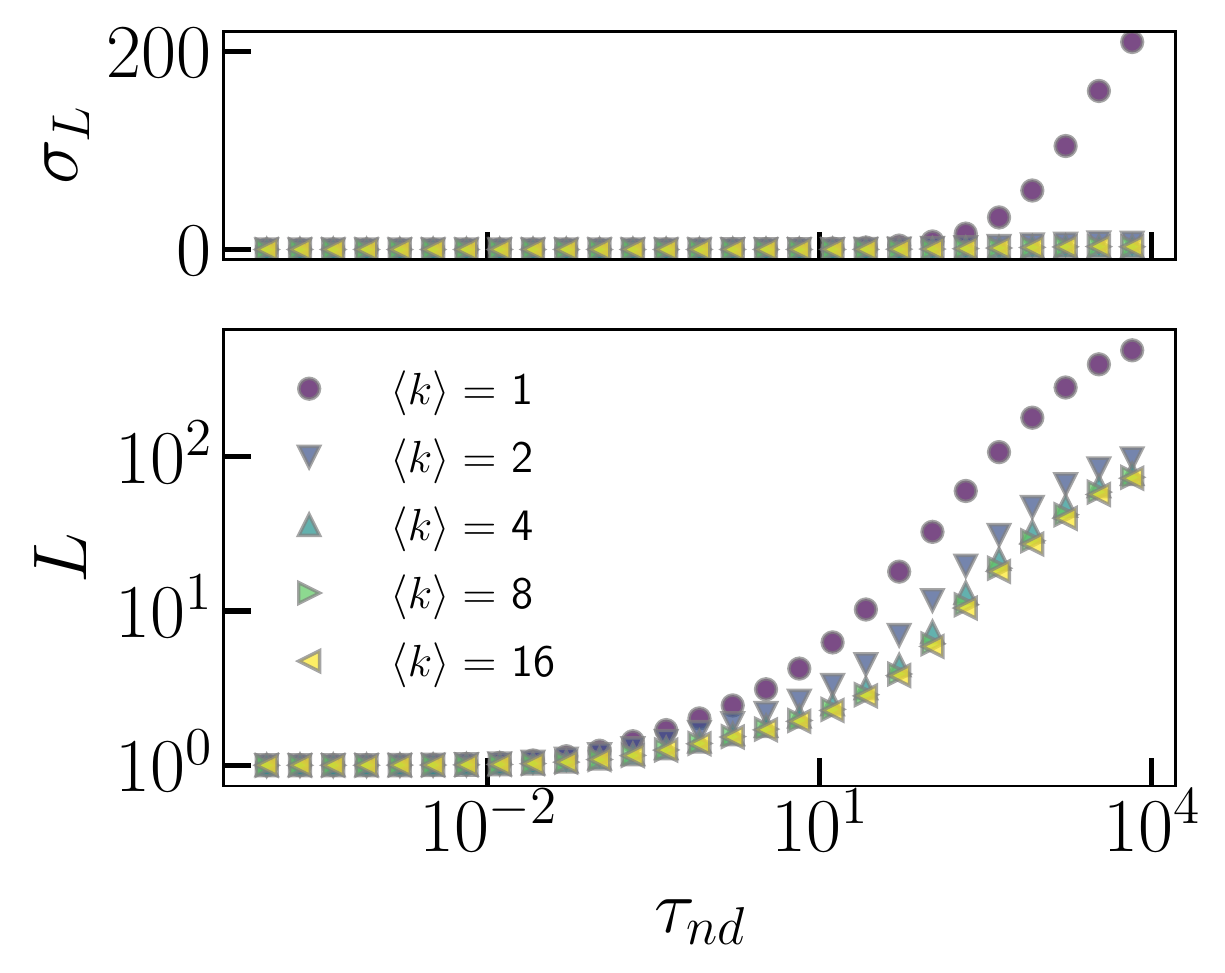}
\includegraphics[width=0.3\textwidth]{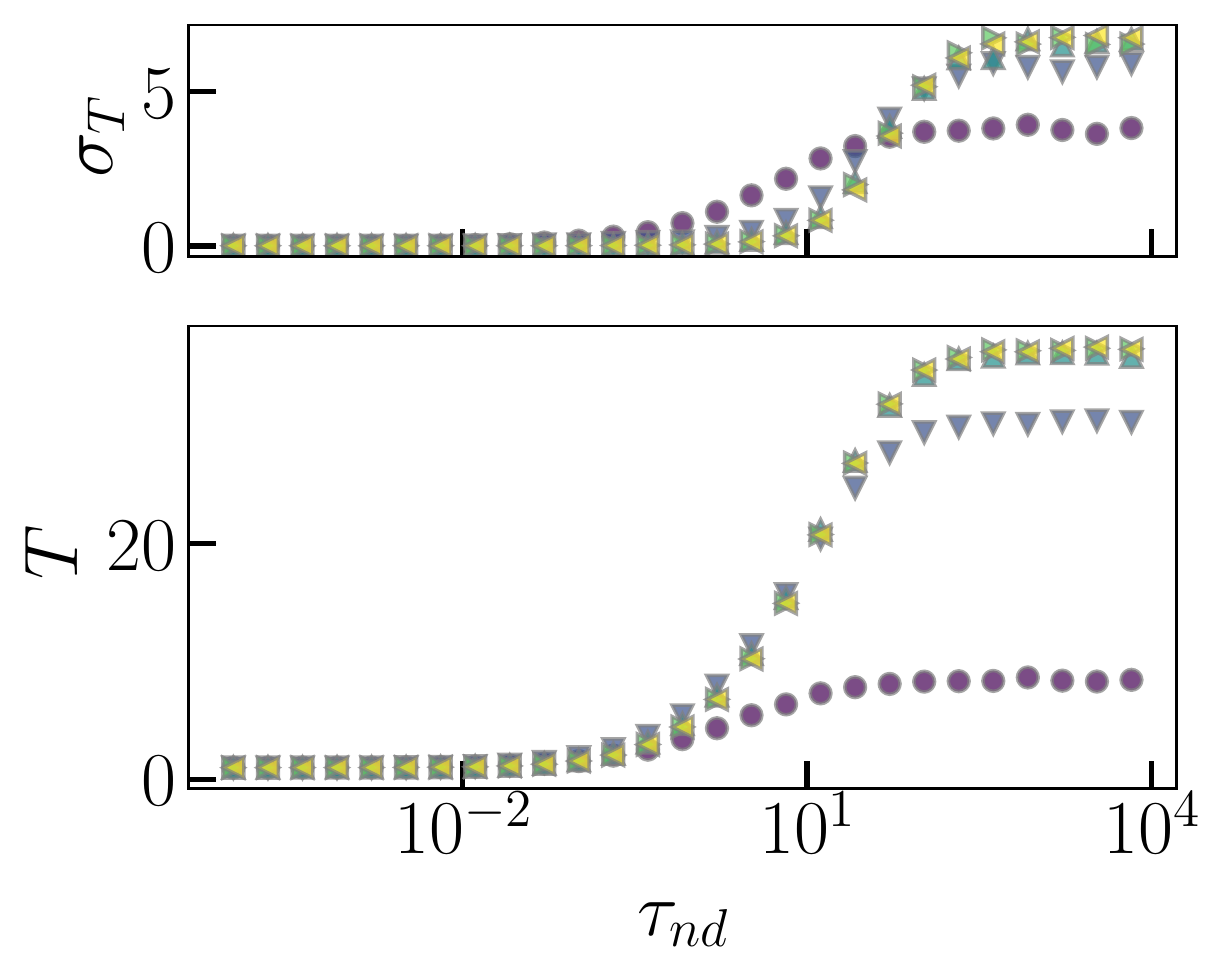}
\end{center}
\caption{\label{fig:4}
Results for the power-law (top) and exponential (bottom) hashing power distributions, with realistic parameters $\alpha=1.5$ and $\lambda=0.05$ for the ER topology as a function of the block delay $\tau_{nd}$. 
Different quantities and the respective standard deviations $\sigma$ are: the mean branch length $L$ and the time $T$ for blocks to be attached to the main chain.
Different colours represent different average degree $\braket{k}$ with fixed number of nodes $N=200$.}
\end{figure}

\begin{figure}[]
\begin{center}
\includegraphics[width=0.3\textwidth]{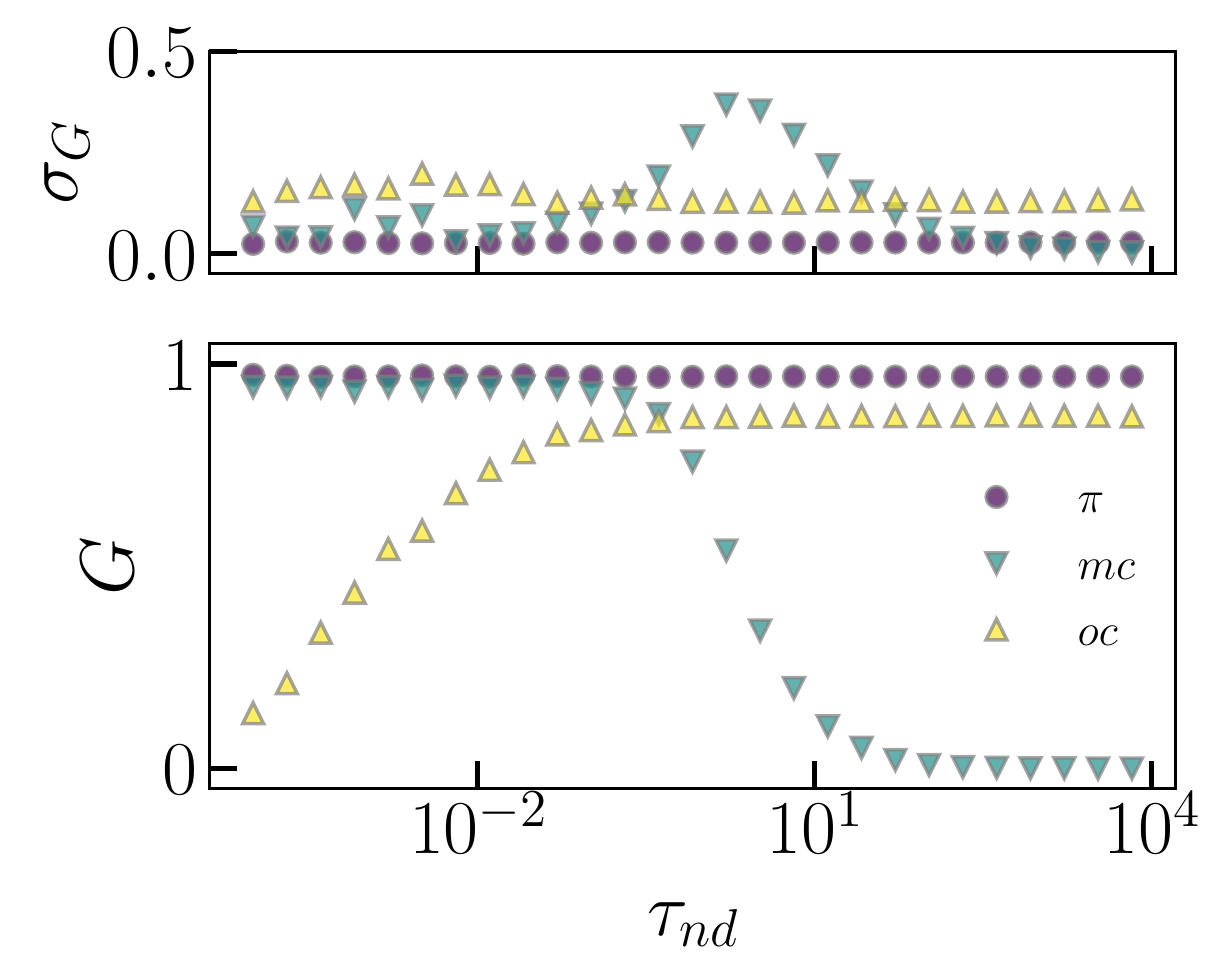}
\includegraphics[width=0.3\textwidth]{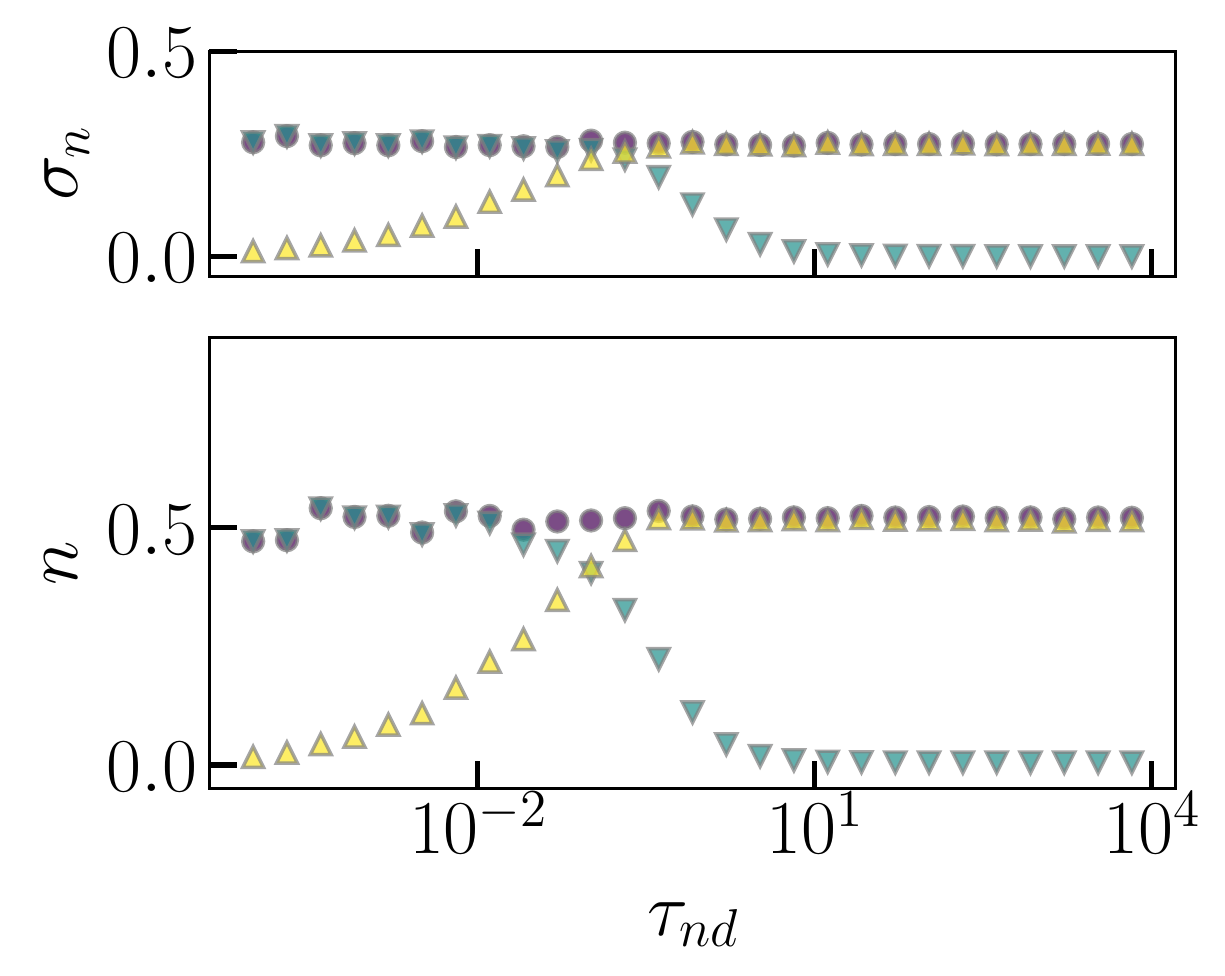}

\includegraphics[width=0.3\textwidth]{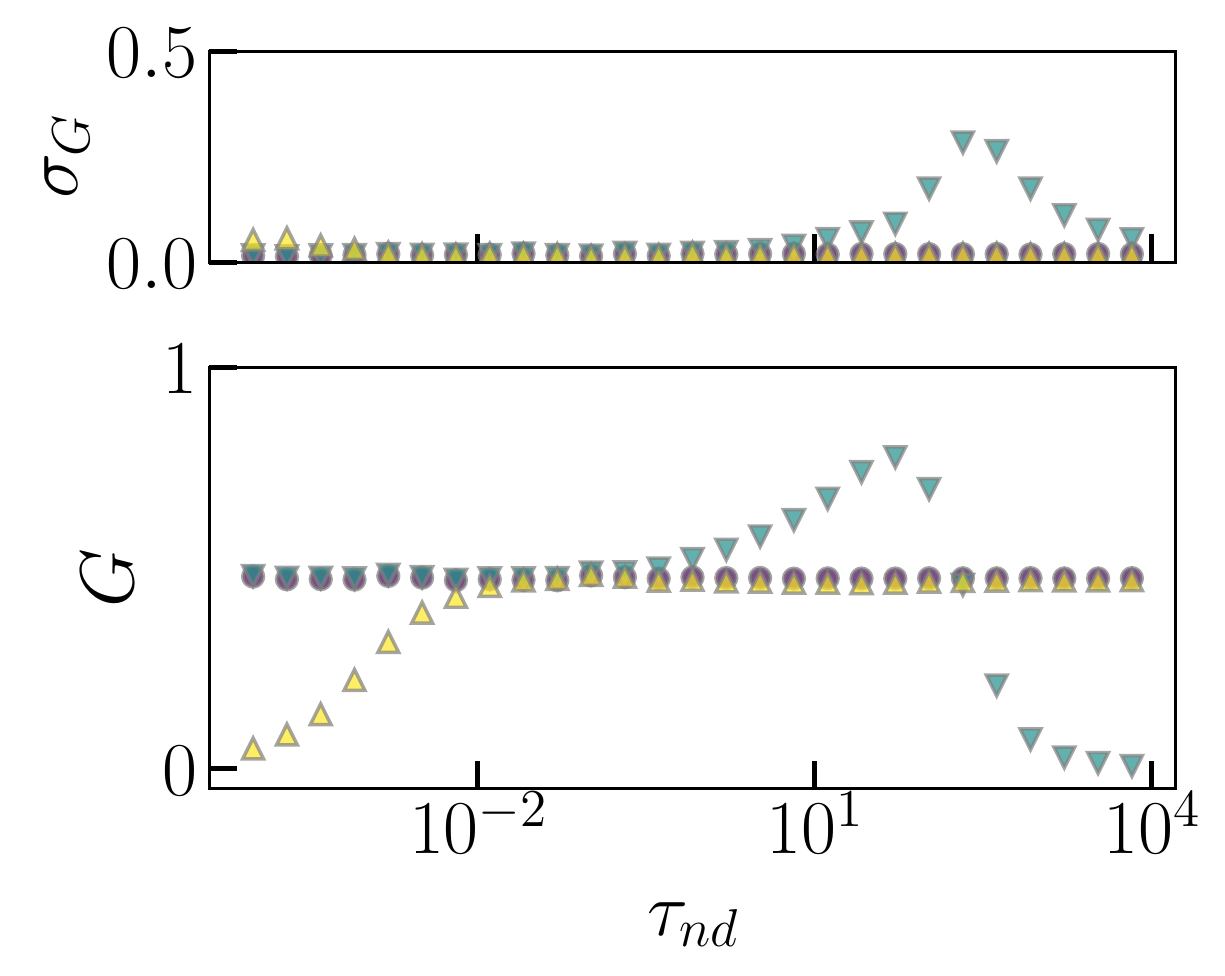}
\includegraphics[width=0.3\textwidth]{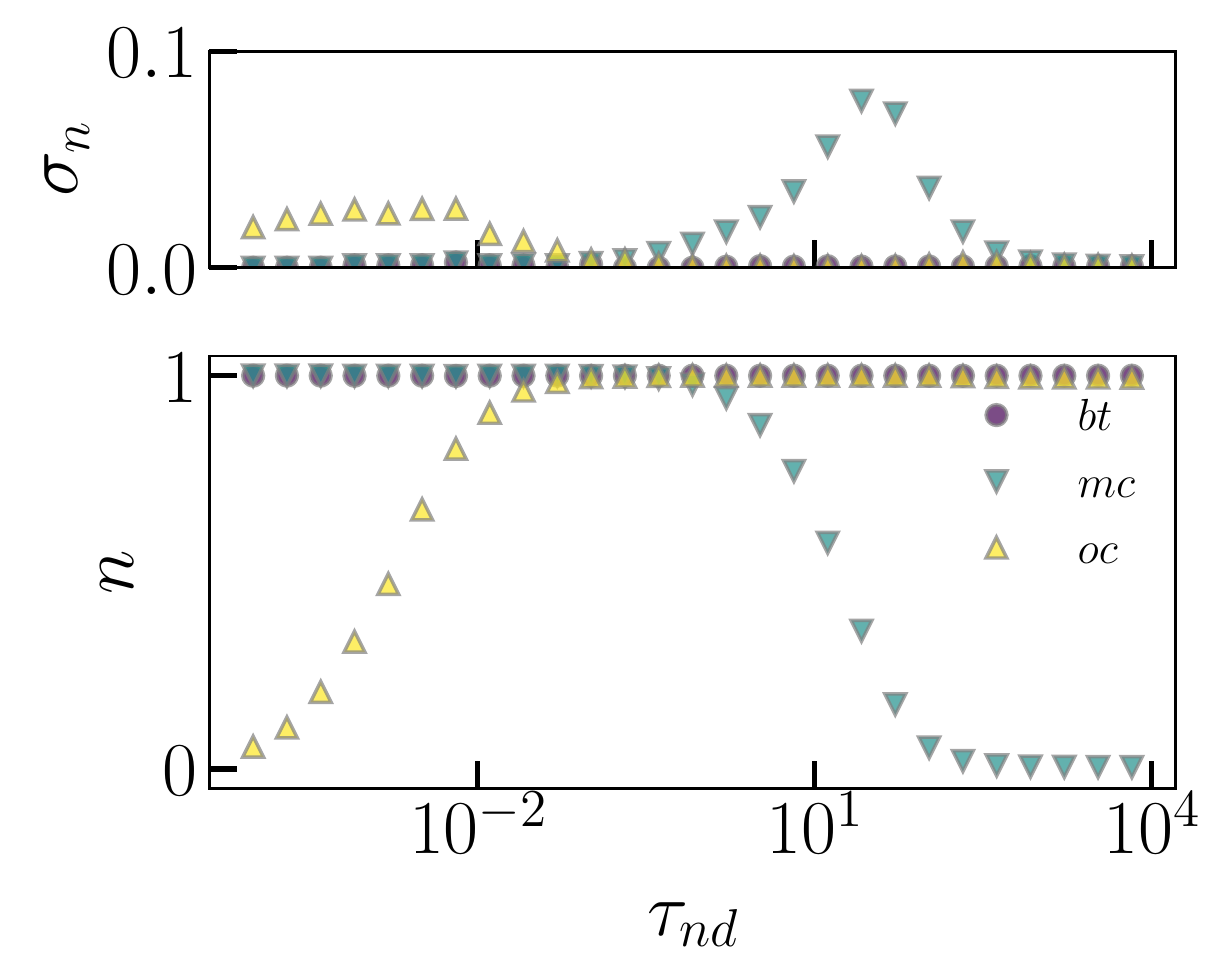}
\end{center}
\caption{\label{fig:5}
Results for the power-law (top) and exponential (bottom) hashing power distributions, with realistic parameters $\alpha=1.5$ and $\lambda=0.05$ for ER topology with $\braket{k}=8$, and $N=200$ nodes each as a function of the block delay $\tau_{nd}$. 
Different quantities and the respective standard deviations $\sigma$ are: the Gini index of the hashing power $G_\pi$, of main-chain blocks $G_{mc}$ and off-chain blocks $G_{oc}$ (left) and the number of miners of blocks in the blocktree $n_{bt}$, in the main chain $n_{mc}$ and off-chain $n_{oc}$ (right).
}
\end{figure}

Blocks in the blocktree are mined by all nodes only for the exponentially distributed hashing powers, while for the power-law case only a fraction of nodes is mining.
For the relative fraction of miners of main-chain and off-chain blocks, $n_{oc}$ grows and saturates for high block delays when eventually all blocks become wasted resources.
Analogously, the fraction of main-chain miners decreases and vanishes for large block delays.

\begin{figure}[]
\begin{center}
\includegraphics[width=0.3\textwidth]{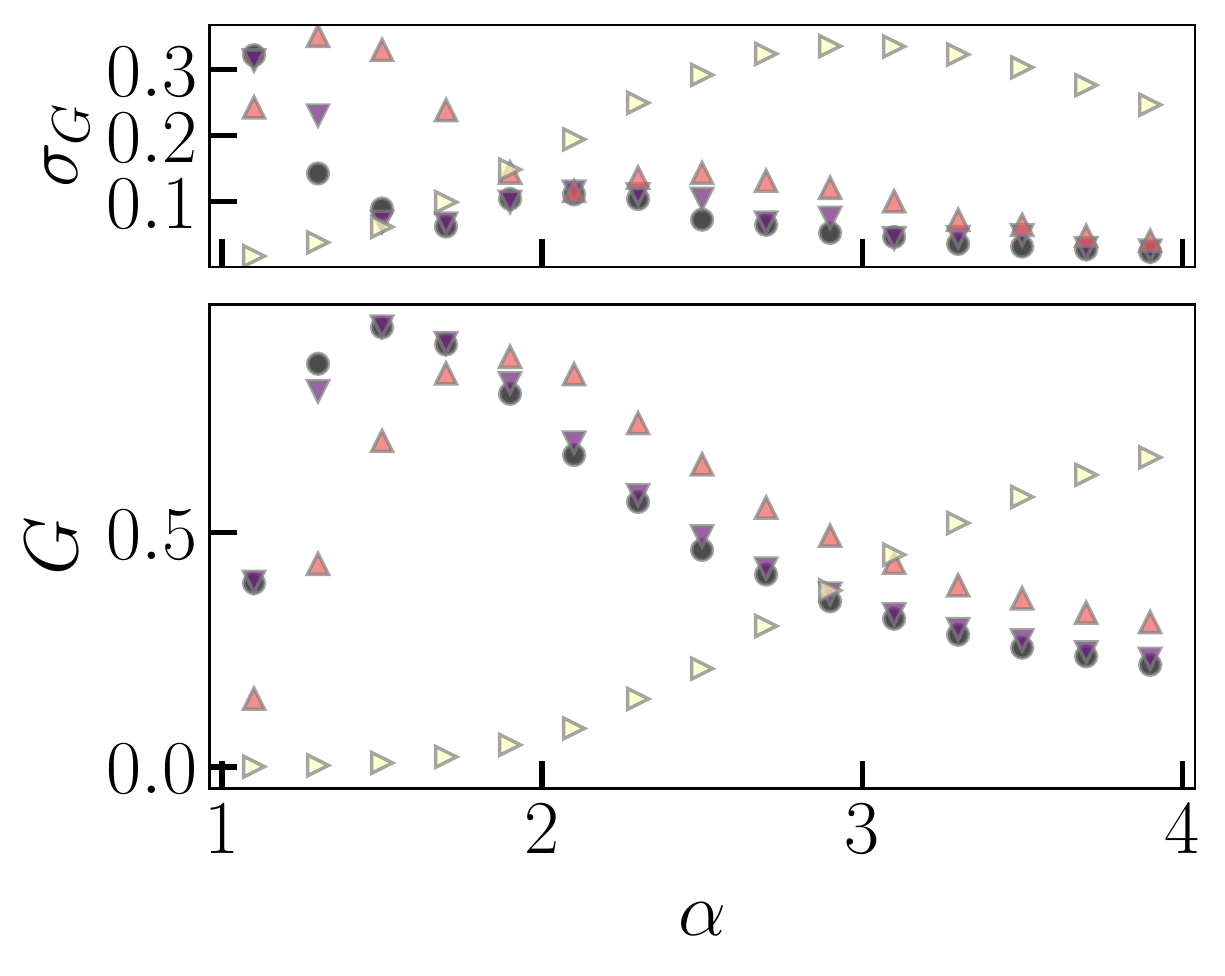}
\includegraphics[width=0.3\textwidth]{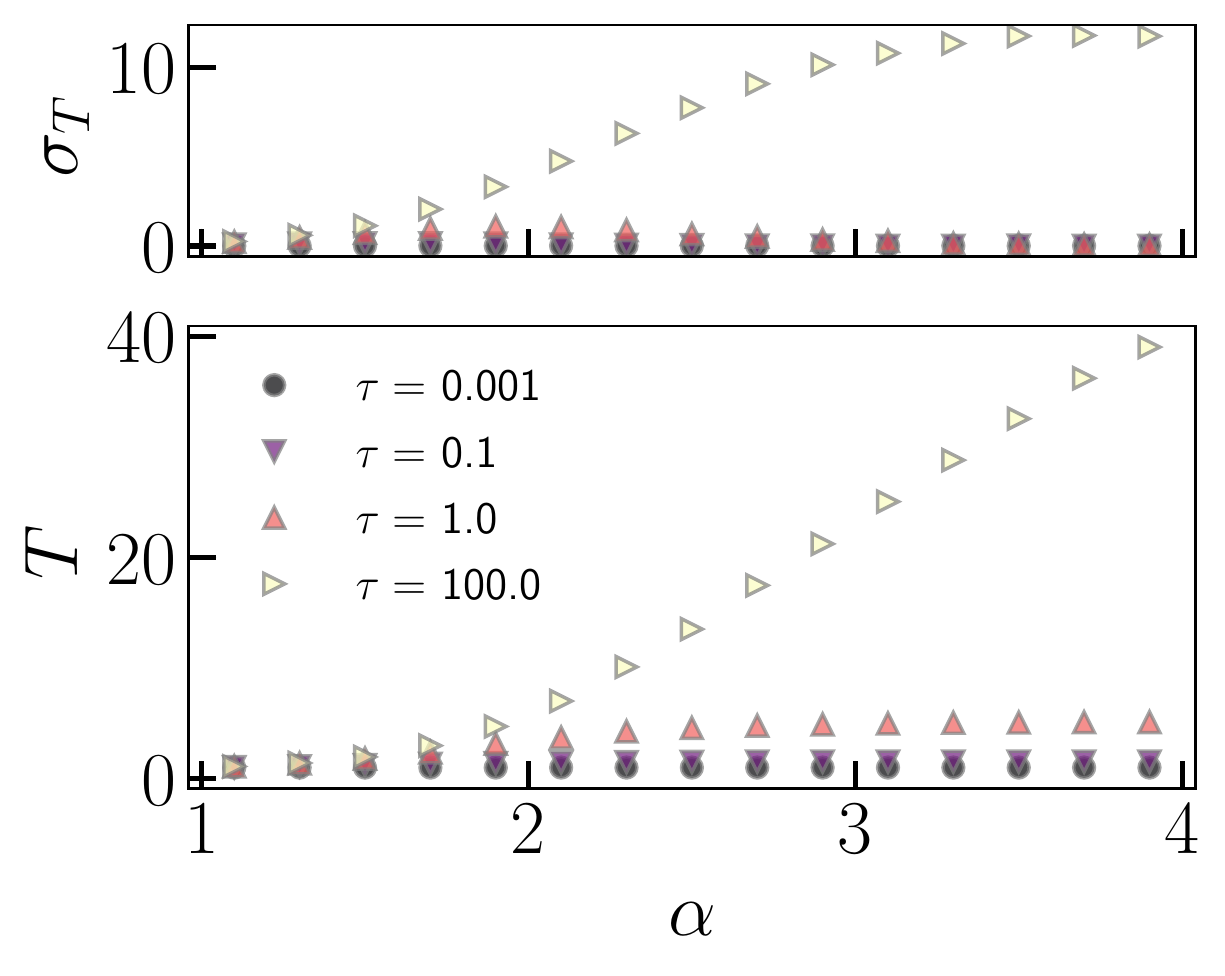}
\includegraphics[width=0.3\textwidth]{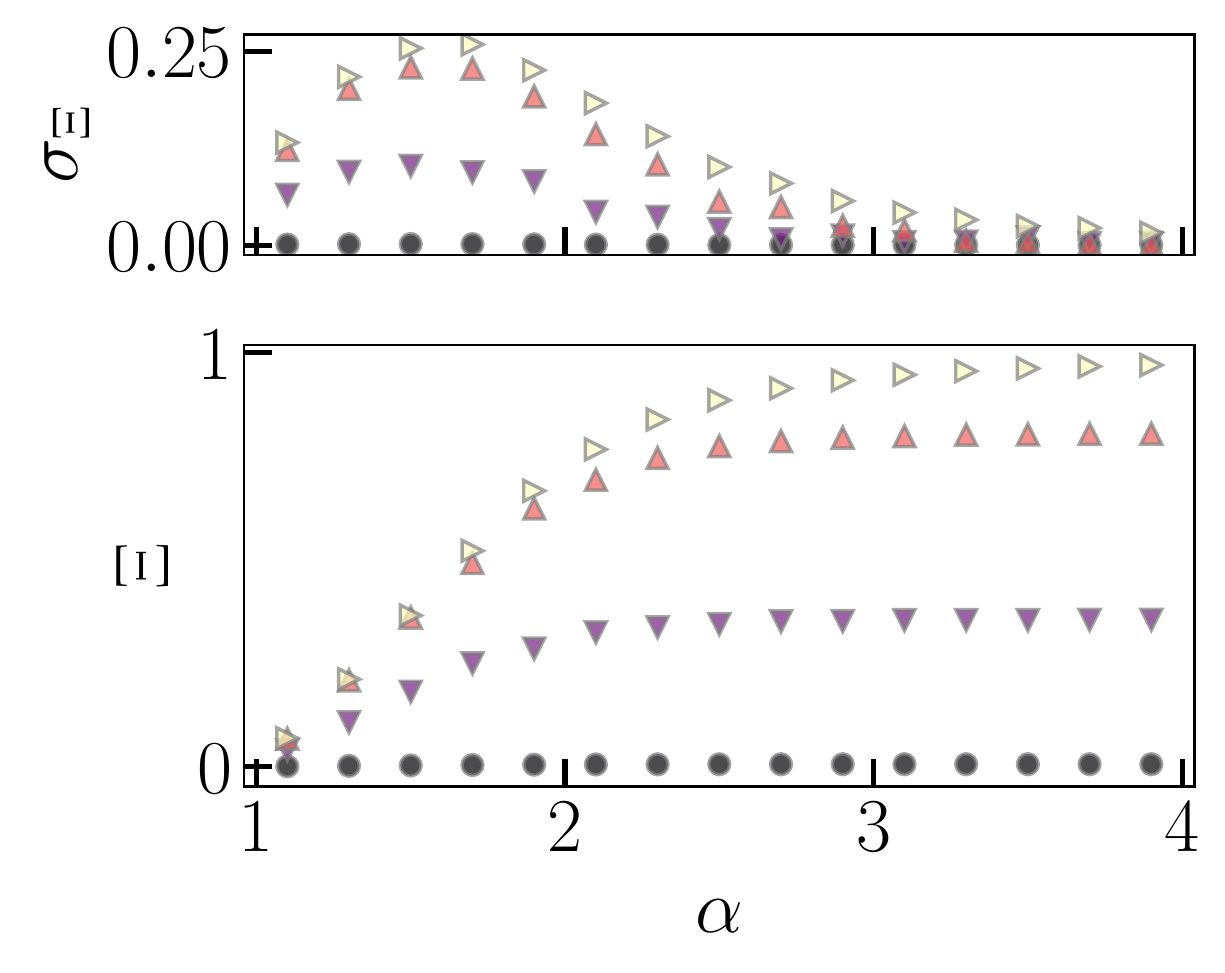}

\includegraphics[width=0.3\textwidth]{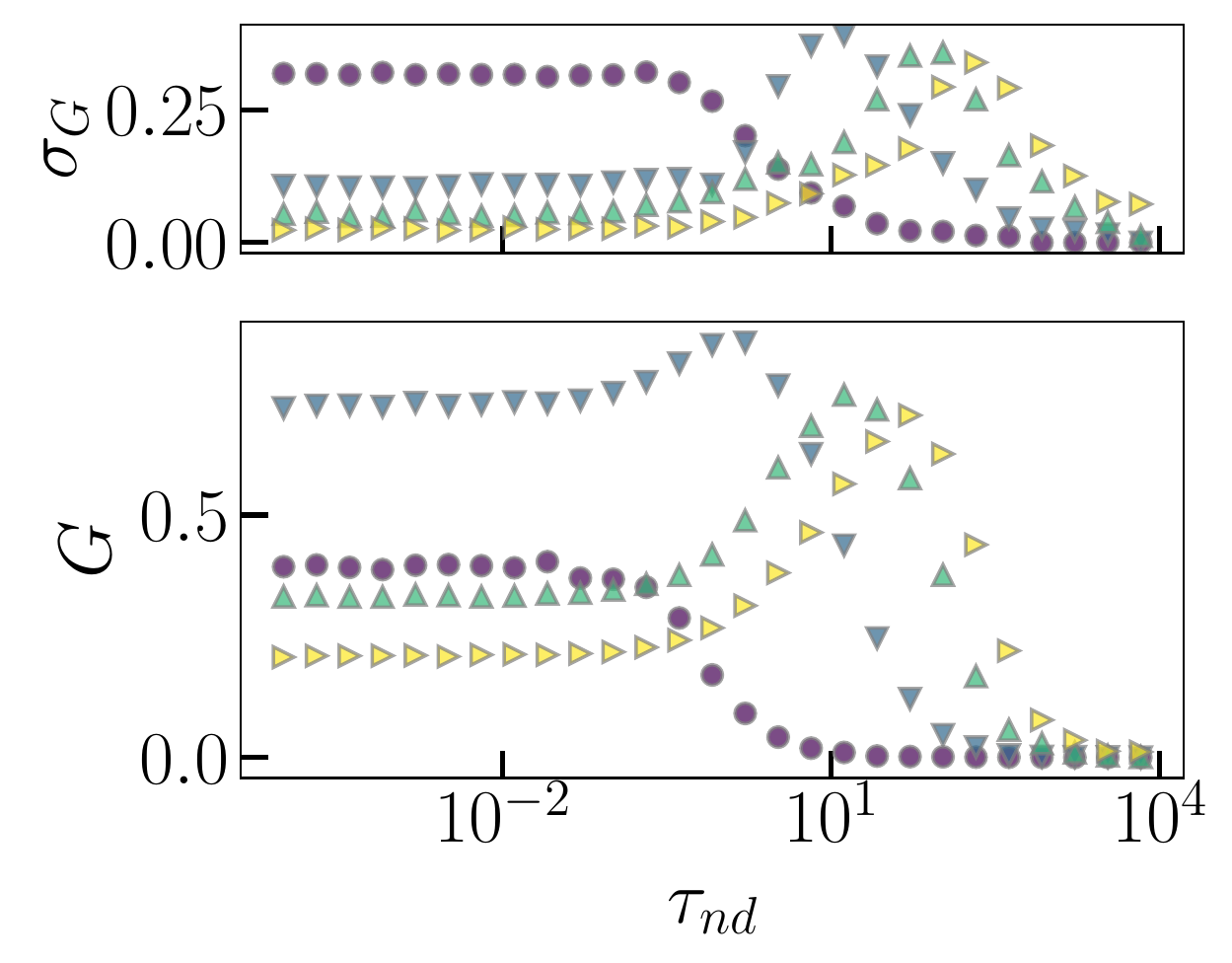}
\includegraphics[width=0.3\textwidth]{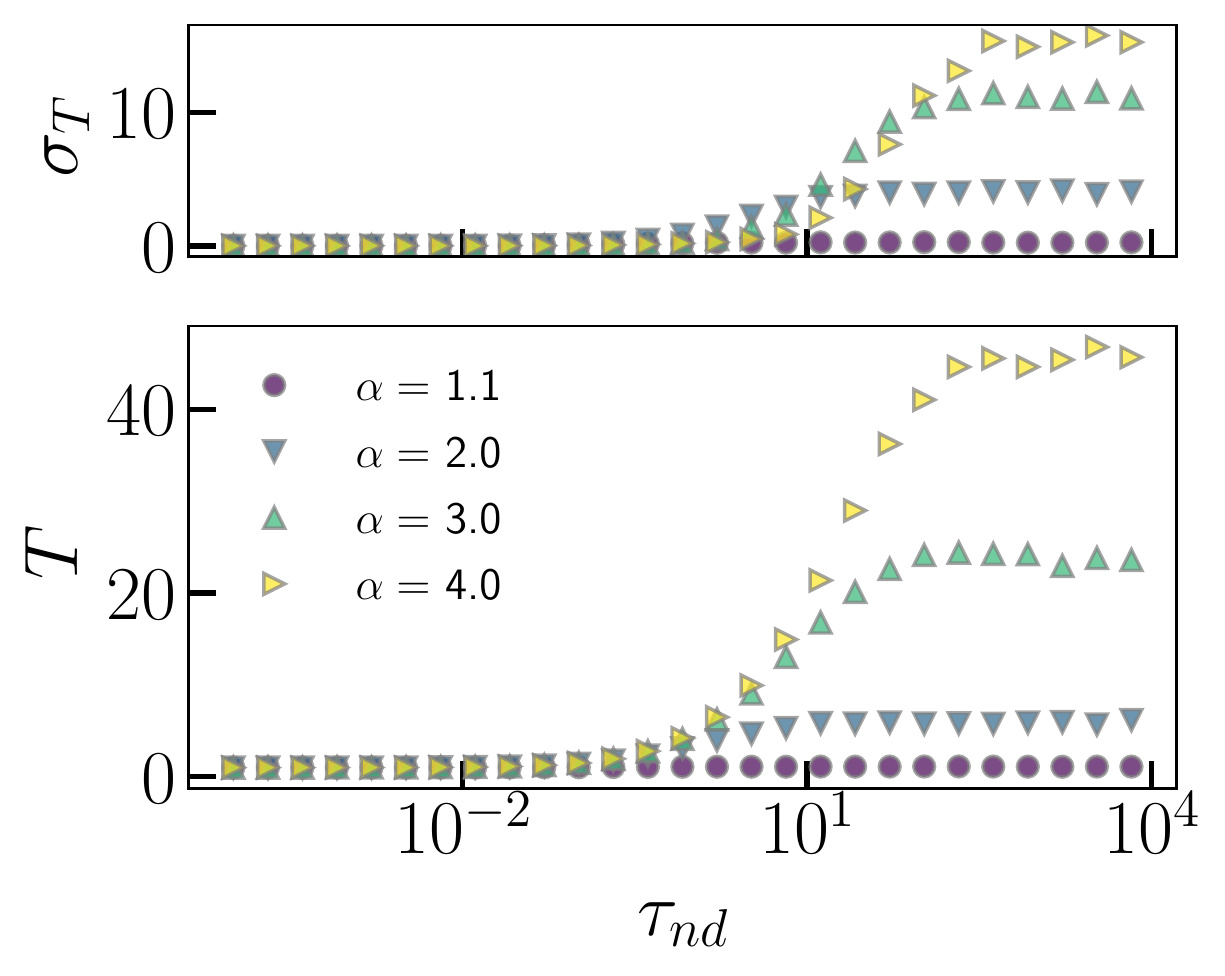}
\includegraphics[width=0.3\textwidth]{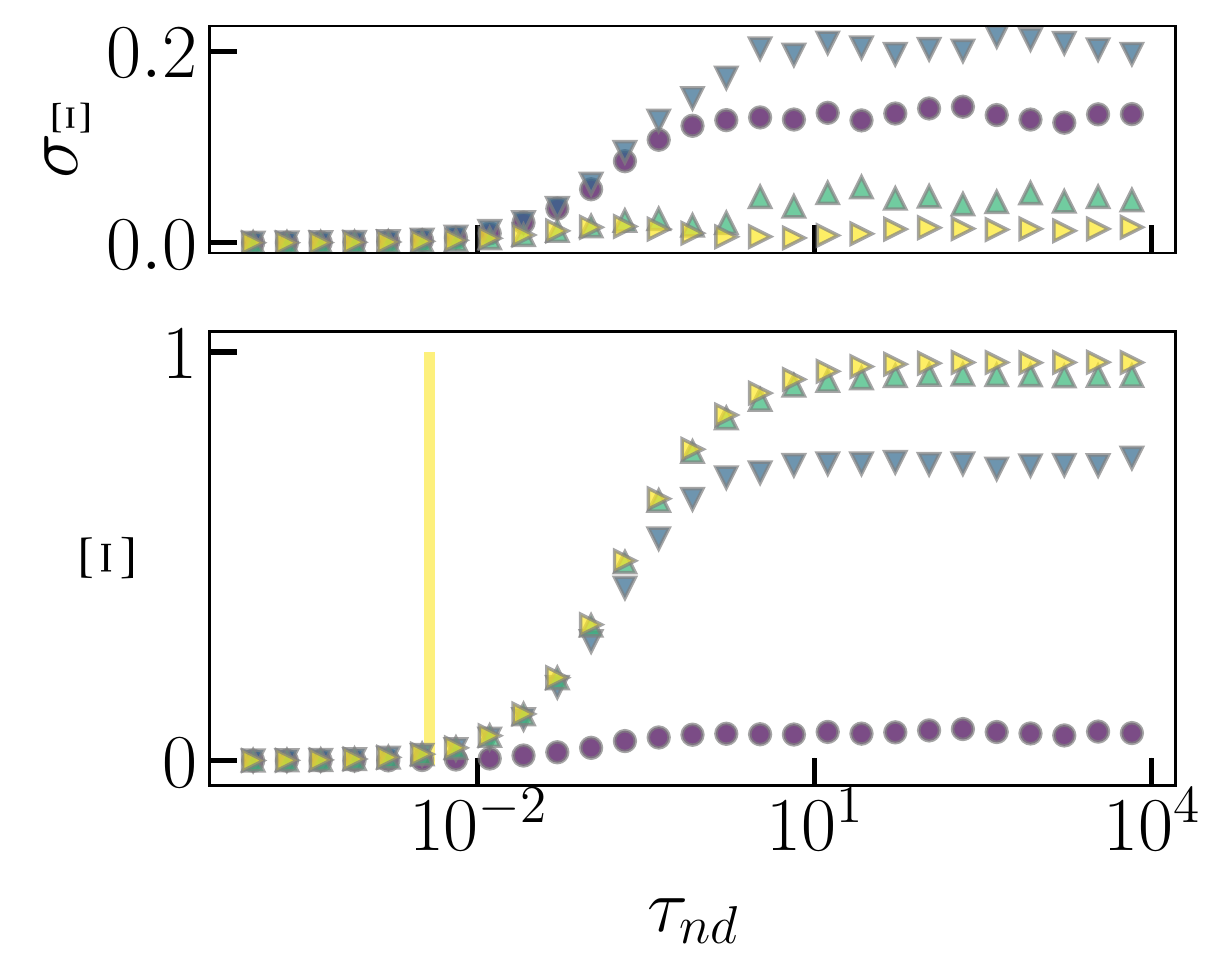}
\end{center}
\caption{\label{fig:6}
Results for ER networks with $k=8$, and $N=200$ with power-law hashing powers. Dependence on the power exponent $\alpha$ for four values of block delay $\tau_{nd}$ (top) and on the block delay for four values of the exponent (bottom). 
Different quantities and the respective standard deviations $\sigma$ are: the Gini coefficient of the main-chain blocks $G$, the time $T$ for blocks to be added to the main chain, and the fraction of orphaned blocks $\Xi$.
}
\end{figure}

To understand how the hashing power heterogeneity changes the system properties and, in particular, how it affects consensus, we focus on the power-law hashing power distribution and study the dependence on the exponent $\alpha$.
In Fig.~\ref{fig:6}, we show for the ER topology with $N=200$ and $\braket{k}=8$ the Gini index of the blocks created in the main chain $G$, the mean propagation time $T$ of blocks in the main chain, and the fraction of orphaned blocks $\Xi$ as a function of the hashing power distribution exponent $\alpha$ for four values of the block diffusion time $\tau_{nd}$ (top panels) and as a function of $\tau_{nd}$ for four values of $\alpha$ (bottom panels).
For low delays (different colours in the same panels) at low exponent $\alpha$ values, the mining process is very heterogeneous and the Gini index for main-chain blocks is maximal within the errors.
Contrary, for large block delays $\tau_{nd}<\tau_b$ there is a lower probability for any node, independently of its hashing power, to mine a block that will be in the main chain.

As $\alpha$ is increased, we increase homogeneity in the mining process, and for low delays we find an increasing number of nodes to be able to mine blocks that belongs to the main chain, see also Fig.~\ref{fig:5}. For $\tau_{nd} = 10>\tau_c>\tau_b$, the system is always not in consensus and a large fraction of blocks, which is higher for more homogeneous hashing powers, is orphaned (right panels) resulting in a very unequal mining process for main chain blocks. 
For the expected time to add blocks in the main chain $T$ (middle panels), when $\tau<\tau_b$, we find no significant dependence on $\alpha$, as most blocks are eventually in the main chain independently of the block miner. For large delays, and as mining becomes more homogeneous, higher $T$ values result. 
These results show that, contrary to the emergent branching process in the blockchain, consensus in this system is a property that is independent of the underlying mining process. 
For $\tau_{nd} \le \tau_b$ the system is partially in consensus and new blocks are consistently attached to the main chain. Then, for $\tau_{nd} > \tau_b$, the time to create blocks in the main chain $T$ (middle panels) and the fraction of orphaned blocks $\Xi$ (right panels) increase the more the mining is homogeneous.
\end{flushleft}

\begin{figure}
\begin{center}
\includegraphics[width=0.3\textwidth]{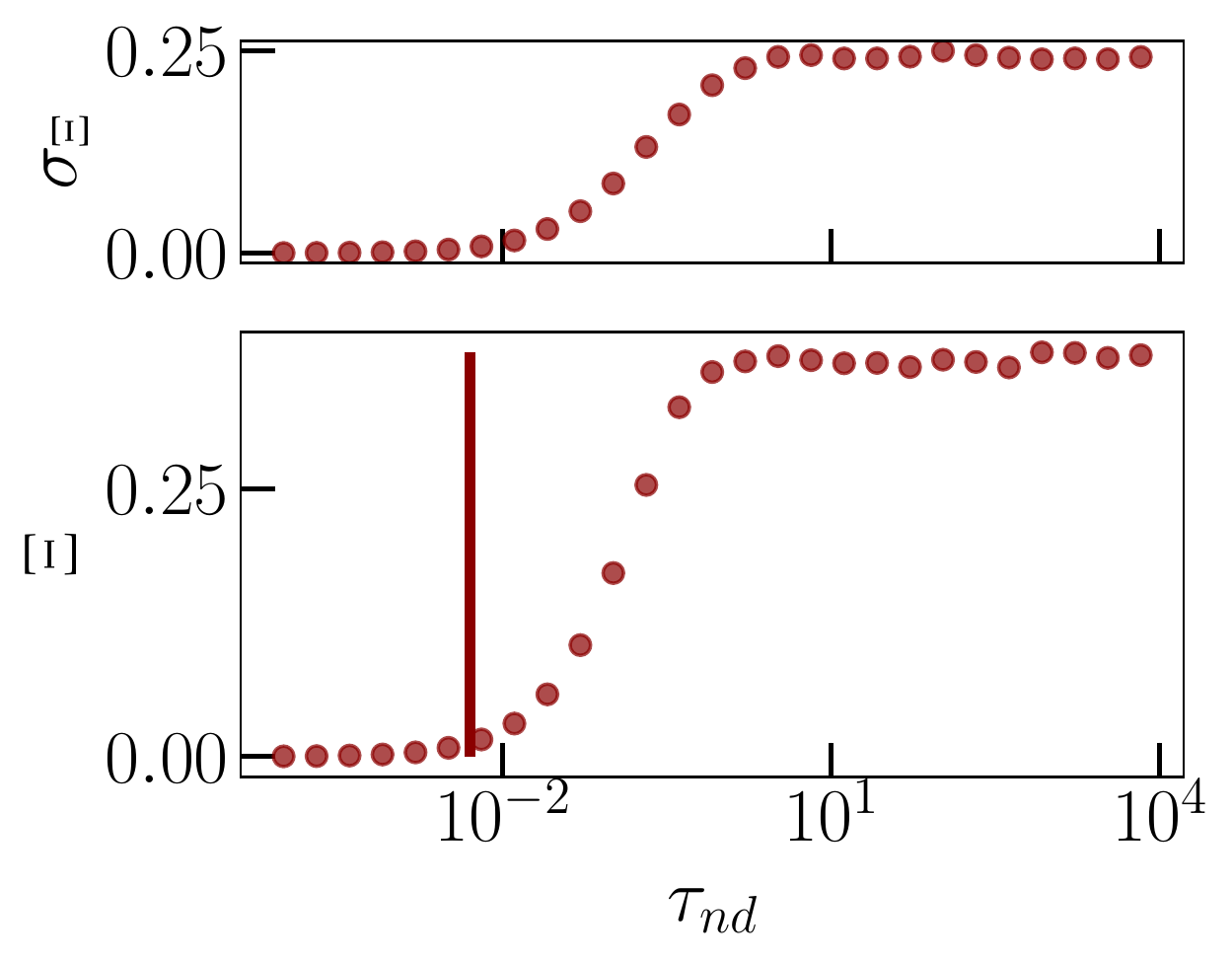}
\includegraphics[width=0.3\textwidth]{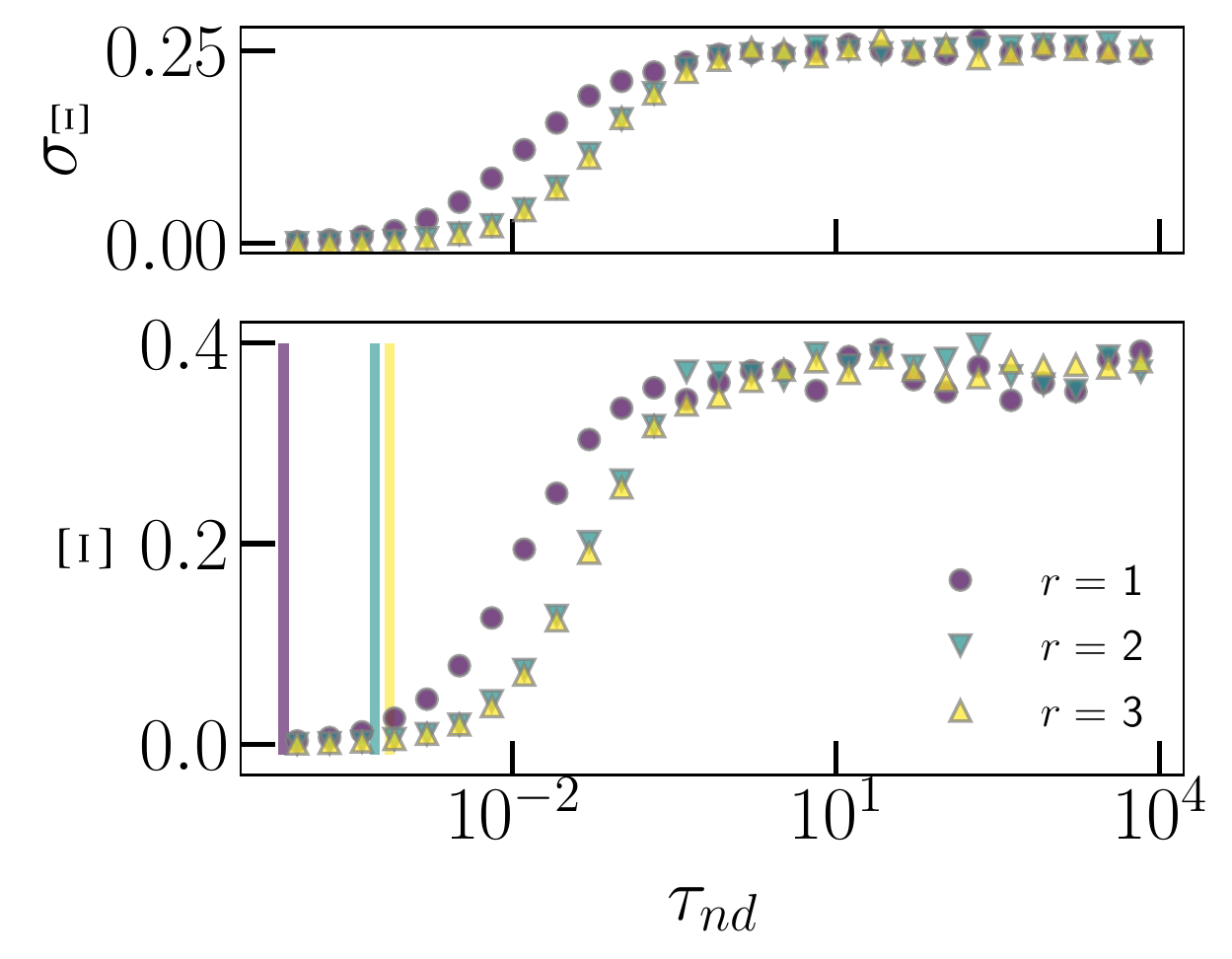}
\end{center}
\caption{\label{fig:7}
Results for the power-law hashing power distribution exponent $\alpha=1.5$ for the complete (left) and tree topology (right) with $N=200$ nodes as a function of the block delay $\tau_{nd}$. 
Different colours in the right panel represent different tree branching factors, and vertical lines mark the estimation of the branching delay $\tau_b=1/\langle M \rangle$.
} 
\end{figure}

In Fig.~\ref{fig:7} we show how the topology of the network and, in particular, the presence of loops can become triggering factors for the emergence of blockchain branches. We compare two opposite scenarios: (i) a complete graph (left panel) with the highest possible number of loops; (ii) a tree graph with different branching factors $r$ (right panel) where loops are absent by construction. 

We find that, complete graphs (Fig.~\ref{fig:7} - left panel) -- having shorter average first passage times -- display a higher branching threshold. In these cases, the system can sustain higher block delays $\tau_{nd}$ without becoming branched.
On the contrary, tree graphs -- having longer average first passage times -- display less efficient block transmission. 
Surprisingly, we also find that, to mitigate this inefficiency, it would be sufficient to increase the branching factor $r$. 
This result suggests a possible strategy to increase the efficiency of the system by increasing communicability in the underlying miners network, e.g. by adding random connections among unconnected miners.

\section{Conclusions}

In this paper, we have introduced a new, minimalistic stochastic model of diffusion processes in P2P networks, that provides novel insights on blockchain consensus mechanisms.

Resorting to simulations and numerical analysis, we were able to identify emergent properties and transition phases that characterise the blockchain consensus when different structural features of the system are changed. 

In particular, we leverage on complex system theory to model the competition in the block creation process among miners in order to uncover the emergence of inefficiencies - i.e., branched blockchains - and detect the conditions under which global consensus is achieved in such decentralised systems. In absence of a competing block creation process, the model reproduces the well-known dynamics of spreading models, where each block diffuses throughout all P2P network and the probability for a non-linear blockchain vanishes, while the system is always in consensus.

Another novelty of our work is the calibration of the model with empirical data collected annually (from 2013 to 2019) about mining activities for three of the most prominent cryptocurrencies in terms of market capitalization ( Bitcoin, Ethereum and Litecoin).
By analysing the annual distributions of hashing power for all three cryptocurrencies, we have compared the two resulting scenarios of heterogeneous and homogeneous distributions, using a power-law and an exponential distribution, respectively. This analysis revealed in each cryptocurrency the more realistic estimation of the distribution parameters to use in our simulations. Among those, we found crucial differences in the block creation shares among the miners specifically between Bitcoin, which is characterised by more homogeneous shares, and Ethereum, where a power-law distribution is systematically a better fit.

The numerical analysis allows us to study the blocktree as the main emergent property, whose topology can be used as a proxy for the efficiency of the system.
In particular, we have analysed how the fraction of orphaned blocks, the ratio of branches and the consensus probability change for different values of the control parameter (the block propagation delay). 
Independently of the heterogeneity of the P2P network, the blockchain becomes forked as the delay in block propagation increases. In addition, we provide an analytical estimation of the characteristic branching block delay $\tau_b$ for different P2P network topologies.

As our main findings, we have showed that --  regardless of the the hashing power distribution of the miners --, the global consensus emerges at a specific block propagation delay which turns to be a key control parameter. 
The analysis of the blocktree structural properties shows that increasing the block delay results in longer branches and higher waiting times for creation of main-chain blocks.

The analysis of the Gini coefficient of mining hashing power of blocks appended to the main chain reveals that the concentration of mining power drops dramatically ($G \to 0$) at increasing levels of network latency. For intermediate latency levels (as observed in real mining networks) we have instead a higher mining concentration ($G \to 1$). 
This holds true for the ER mining network both under its power-law and exponential characterization.
By excluding the (unrealistic) boundary parameters levels, interestingly we highlight that for ER mining networks with power-law distribution there exist an optimal intermediate pair of network latency and node heterogeneity that minimize the coefficient of mining inequality $G$. 

Finally, we have showed that -- for different amounts of delay in block propagation --  blockchain branching heavily depends on the network topology of the mining network. 

While minimalistic, our model is easily extensible to encompass real-world information such as inherent network delays, changing conditions, or specific node properties.
The stochastic modelling approach to the consensus protocol in blockchain systems introduced in this paper can also be helpful for redesigning of protocols for network formation in such systems and to devise possible incentive schemes to increase the system efficiency.


\section*{Acknowledgments:}    FI acknowledges funding of SNSF (Switzerland) through project  200021\_182659. CJT acknowledges financial support from the  University of Zurich through the University Research Priority Program on Social Networks. PT acknowledges support from UBRI Connect.

\section*{Author contributions statement}

CJT and PT  conceived the experiment.  PT conceived an initial version of the model and CJT extended the development and performed the numerical simulations. FI  analysed the empirical data and introduced the theoretical approach.  
All authors wrote the   manuscript,  polished and approved the final version. 

\bibliography{literature}

\end{document}